\documentclass[twocolumn]{aa-package/aa}
\usepackage{graphicx,color}
\usepackage[varg]{txfonts}
\usepackage{natbib}
\usepackage[font={small,rm}]{caption}
\usepackage{array}
\usepackage{bm}
\usepackage{color}
\usepackage[T1]{fontenc}
\usepackage{multirow}
\usepackage{caption}
\usepackage{footnote}

\usepackage{xcolor}

\begin{document}

\title{A model exploration of NIR ro-vibrational CO emission as a tracer of inner cavities in protoplanetary disks}


 

\author{Antonellini, S.$^{1,2}$\and Banzatti, A.$^{3}$\and Kamp, I.$^{1}$\and Thi W.-F.$^{4}$\and Woitke, P.$^{5}$}
\authorrunning{Antonellini et al.}

\institute{Kapteyn Astronomical Institute, Postbus 800, 9700 AV Groningen, The Netherlands\\
\email{antonellini@astro.rug.nl}\and 
Astrophysics Research Centre, School of Mathematics and Physics, Queens University Belfast, Belfast BT7 1NN, UK\and 
Lunar and Planetary Laboratory, The University of Arizona, Tucson, AZ85721, USA\and 
Max-Planck-Institut fu\"{r} extraterrestrische Physisk, Giessenbachstrasse 1, 85748 Garching, Germany\and
SUPA, School of Physics \& Astronomy, University of St. Andrews, North Haugh, St. Andrews KY16 9SS, UK}

\abstract
 {Near-IR observations of protoplanetary disks provide information about the properties of the inner disk. High resolution spectra of abundant molecules such as CO can be used to determine the disk structure in the warm inner parts. The $v2/v1$ ro-vibrational ratio of $v_{1-0}$ and $v_{2-1}$ transitions has been recently observed to follow distinct trends with the CO emitting radius, in a sample of TTauri and Herbig disks; these trends have been empirically interpreted as due to inner disk depletion from gas and dust.}
 {In this work we use existing thermo-chemical disk models to explore the interpretation of these observed trends in ro-vibrational CO emission.}
 {We use the radiation thermo-chemical code ProDiMo, exploring a set of previously published models with different disk properties and varying one parameter at a time: the inner radius, the dust-to-gas mass ratio, the gas mass. In addition, we use models where we change the surface density power law index, and employ a larger set of CO ro-vibrational levels, including also fluorescence from the first electronic state. We investigate these models for both TTauri and Herbig star disks. Finally, we include a set of DIANA models for individual TTauri and Herbig disks which were constructed to reproduce a large set of multi-wavelength observations.}
 {This modeling exploration highlights promising parameters to explain the trends observed in ro-vibrational CO emission. Our models with an increasing inner radius can naturally match the trend observed for TTauris, where the vertical spread in the data could also be accounted for by different values for dust-to-gas mass ratio and disk gas mass in different disks. Our models match the CO vibrational ratio observed in Herbig disks only in the case of large inner holes, instead, and cannot produce the low ratios measured in many disks. The models do produce an inversion in the trend, where $v_{2-1}/v_{1-0}$ increases instead of decreasing for CO radii larger than a few au. This is a consequence of the P(4) $v_{2-1}$ line becoming optically thin and super-thermally excited. In our models, this does not require invoking UV fluorescence pumping.} 
 {Our modeling explorations suggest that the observed decrease of $v_{2-1}/v_{1-0}$ with CO radius in TTauri disks can be a consequence of inside-out disk depletion. For the Herbig disks, instead, a more complex inner disk structure may be needed to explain the observed trends in the excitation of CO emission as a function of emitting radius: disk gaps emptied of dust, partially depleted in gas and/or possibly a disk structure with inverted surface density profile. These structures need to be further investigated in future work.
} 


\maketitle

\section{Introduction} \label{sec:intro}
Disks around young stars evolve and disperse on timescales of a few Myr \citep[e.g.][]{evans2,hernandez,fedele3,dent1}. Several processes can drive this evolution, including disk winds, planet formation, dust growth and migration, and not all disks necessarily undergo the same path of evolution \citep[e.g.][]{alexander,testi}. When assuming an exponential decrease with time of the dust mass (e-folding), the expected disk lifetime is around 2-3 Myr for the inner disk and 4-6 Myr for the outer disk, as inferred from  near- and mid-IR continuum observations \citep{ribas}. This has been long discussed within clearing scenarios produced by winds or planet formation \citep[e.g.][]{Owen16,ercolano}.
Spectral Energy Distributions and recently also interferometry and direct imaging show the presence of inner gaps and/or holes in protoplanetary disks \citep[see e.g.][for a recent observational review]{espaillat}. Understanding the nature of these gaps/holes in terms of their dust and gas content is important in deciphering the underlying process responsible for the inner disk (<10-30 au) evolution, the main region of planet 
formation.
Atomic and molecular line profiles can be effectively used as disk gas depletion indicators, if $M_\mathrm{star}$ and the disk inclination are known. Some transition disks (based on SED classification) have line detections for CO in the infrared, H$_2$ in the ultraviolet, or [OI] in the optical; these gas tracers can be used as important diagnostics for inner disk structure and evolution \citep[e.g.][]{salyk3,pontopp11,bp15,simonM16,hoadley,banz19}. 

The CO ro-vibrational lines emitting at $\sim$4.6-5 $\mu$m are key diagnostics of this inner disk region, as they typically need the warm and dense molecular gas within a few au from the star to be excited 
\citep[e.g.]{brittain,najita3,blake}. High spectral resolution line profiles provide information on CO excitation and emitting disk radii from the observed gas kinematics \citep[e.g.][]{brittain2,salyk4,vanderplas,bertelsen}. 
Recently, a new analysis of high velocity resolution CO ro-vibrational spectra from a large sample of T~Tauri and Herbig disks defined a temperature-radius (T-R) diagram, by finding distinct trends in CO excitation as a function of emitting radius \citep{bp15}. When emitting from $\lesssim 1$\,au, CO lines show a decreasing vibrational excitation with increasing emitting radius, which was interpreted as due to IR pumping of CO following the temperature profile of the local warm dust. 
When emitting from $\gtrsim 1$\,au, instead, CO lines show an inversion in the temperature profile (that in this work we call upturn), with the vibrational excitation increasing with disk radius, interpreted as UV fluorescence in cold CO gas located beyond an inner disk cavity. 
This second trend in the T-R digram, in particular, is populated by CO spectra observed in Herbig disks, while TTauri disks populate the first trend. The vibrational CO sequence revealed in the CO temperature-radius diagram was discussed by 
\citet{bp15} and \citet{banz17,banz18} in the context of a gap opening scenario, where CO emission recedes to larger disk radii together with dust depletion possibly in an inside-out fashion.

Some recent modeling from \citet{woitke4} with the code ProDiMo, showed that the line flux from $v_{1-0}$ R(10) is indeed affected by moving the inner radius of disks, and in particular, due to a combination of line emitting region, and weaker FUV stellar flux, the rovibrational line flux increases until $R_\mathrm{in}=3~$au surface area and then it starts to decrease due to loss of excitation \citep[Fig.19 in][]{woitke4}. 
\noindent
They also found dust properties and disk flaring to affect the line flux of the $v_{1-0}$ rovibrational lines.
In a recent work, \citet{bosman} performed slab modeling and detailed DALI simulations to explain the ro-vibrational ratios observed towards Herbig star disks published in \citet{banz17,banz18}, using a CO model molecule with five vibrational levels, each with 40 rotational states. From their slab model studies of the CO ro-vibrational lines they concluded that the vibrational ratios of $\sim\!0.2$ observed for disks with large cavities can be reached with high column densities and low temperatures. Their thermo-chemical disk geometries cannot naturally reproduce the ''upturn'' observed in the data \citep{bp15}. However, their exploitation of the parameter space leads them to conclude that high ro-vibrational ratios ($v_2/v_1~>~0.2$) at CO emitting radii larger than 5~au can be produced using gas-to-dust mass ratios larger than 10$^4$. Such high values can for example be associated with the formation of dust traps. 
The CO T-R diagram has been so far analyzed in terms of the empirical correlations found in the data. 
In this work, we have compared (mostly) previously existing models to the CO sequence measured in the T-R diagram by \citet{bp15}, with the main goal of testing and improving the interpretation of the observed trends in CO ro-vibrational emission.

The approach we follow in this paper is to use previous models that explored fully devoid disk holes of increasing size \citep{antonellini, antonellini1}, and extract from our models parameters that can be compared to the observational analysis from \citet{bp15}. In addition, we include a number of existing disk models for individual sources taken from the DIANA project \citep{kamp5,woitke5}.
In Section~\ref{sec:mod}, we present our modeling and our method of comparison with the observations. In Section~\ref{sec:res}, we compare our model prediction with the observations. Finally, in Section~\ref{sec:d&c}, we present our interpretation of the CO rovibrational sequence, and the observed upturn in the ro-vibrational ratio vs $R_\mathrm{CO}$ trend. 

\section{Modeling}\label{sec:mod}

\subsection{Disk model series}

The models used for comparison to the data are made with the radiation thermo-chemical code ProDiMo \citep{woitke}. In particular, we look at the CO ro-vibrational lines predicted from a series of previously published models for TTauri and Herbig disks (Table~\ref{Overview}) with different inner radii ($R_\mathrm{in}$), disk gas masses ($M_\mathrm{gas}$, keeping $M_\mathrm{dust}$ fixed), dust-to-gas mass ratios (d/g, keeping $M_\mathrm{gas}$ fixed) as described in \citet[][]{antonellini,antonellini1}. 

We included two additonal unpublished model series with 
different surface density power law index ($\epsilon$), and four others with a larger number of rotational, vibrational and electronic energy levels of CO (Table~\ref{Enea}), needed to test the excitation mechanism responsible for the observed trends. In addition, we included six Herbig disk models combining parameters as $M_\mathrm{gas}$ and $R_\mathrm{in}$, dust-to-gas mass ratio and the elemental abundance of carbon ($\delta_\mathrm{C}$) in order to further test our conclusions. The carbon abundance is defined as 
\begin{equation}\label{CosmicAb}
    \delta_\mathrm{X}=\mathrm{Log}\left(\frac{n_\mathrm{X}}{n_\mathrm{H}}\right)+12\,\,\, ,
\end{equation}
where $n_\mathrm{X}$ and $n_\mathrm{H}$ are the densities of the species X and hydrogen.
In these ``combined models" we change two parameters at the same time with respect to the standard model, in order to investigate their combined effects on CO emission lines (Table~\ref{Overview}, series 5 and 6).

\begin{table*}
\caption{Disk model series used in this study}
\centering
\begin{tabular}{c|c|c|c|c}\hline\hline
Series & Parameter varied & TTauri values & Herbig values & Other changing parameters\\ \hline
(1) & $R_\mathrm{in}$ [au] & 0.05, $\bm{0.1}$, 0.5, 1.0, 5.0, 10.0, 15.0 & 0.1, $\bm{0.365}$, 0.5, 1.0, 5.0, 10.0, & - \\
& & & 15.0, 20.0, 25.0, 30.0 &  \\
(2) & $\epsilon$ & -1.5, -1.0, -0.5, 0.0, 0.5, & -1.5, -1.0, -0.5, & - \\
 & & 0.75, $\bm{1.0}$, 1.25, 1.75 & -0.25, 0.0, $\bm{1.0}$, 1.5 & \\
(3) & $M_\mathrm{gas}$ [$M_{\odot}$] & 0.00001, 0.001, $\bm{0.01}$, 0.05, 0.1 & 0.00001, 0.001, $\bm{0.01}$, 0.05, 0.1 & d/g  \\
(4) & d/g & 0.001, $\bm{0.01}$, 0.1, 1.0, 10.0, 100.0 & 0.001, $\bm{0.01}$, 0.1, 1.0, 10.0, 100.0 & $M_\mathrm{disk}$ \\
(5) & $R_\mathrm{in}$+$M_\mathrm{gas}$ [au,~$M_\mathrm{\odot}$] & &
1.0~\&~0.001 (a), 5.0~\&~0.001 (b) & d/g\\
(6) & $d/g$+$\delta_\mathrm{C}$ & &
0.1~\&~7.08 (c), 0.1~\&~6.08 (d), & -\\
& & & 1.0~\&~7.08 (e), 1.0~\&~6.08 (f) & \\
\hline
\end{tabular}
\tablefoot{Bold font numbers are values for the standard model. The only model series that uses the large CO molecule is series 1}
\label{Overview}
\end{table*}

\subsection{Line fluxes}
\label{2.2}

The non-LTE level populations for the CO molecule are derived from two-dimensional escape probability. Subsequently, the vertical escape line flux is calculated for each transition to explore a large set of disk models. The standard CO ro-vibrational model molecule used (hereafter called ''small CO molecule''), include the first three vibrational levels, and 50 rotational levels for each vibrational state (Table~\ref{Enea}). A complete description of the CO ro-vibrational 	model molecule in ProDiMo can be found in~\citet{thi13}.		
Considering that our nominal disk model is inclined by 30$\degr$ and that CO ro-vibrational lines are emitted from the innermost disk regions with a relevant contribution from the inner wall, a more detailed radiative transfer is used for the series 1 of Table~\ref{Overview} to verify our results, as explained in the next paragraph.

\begin{figure}
\includegraphics[width=0.5\textwidth]{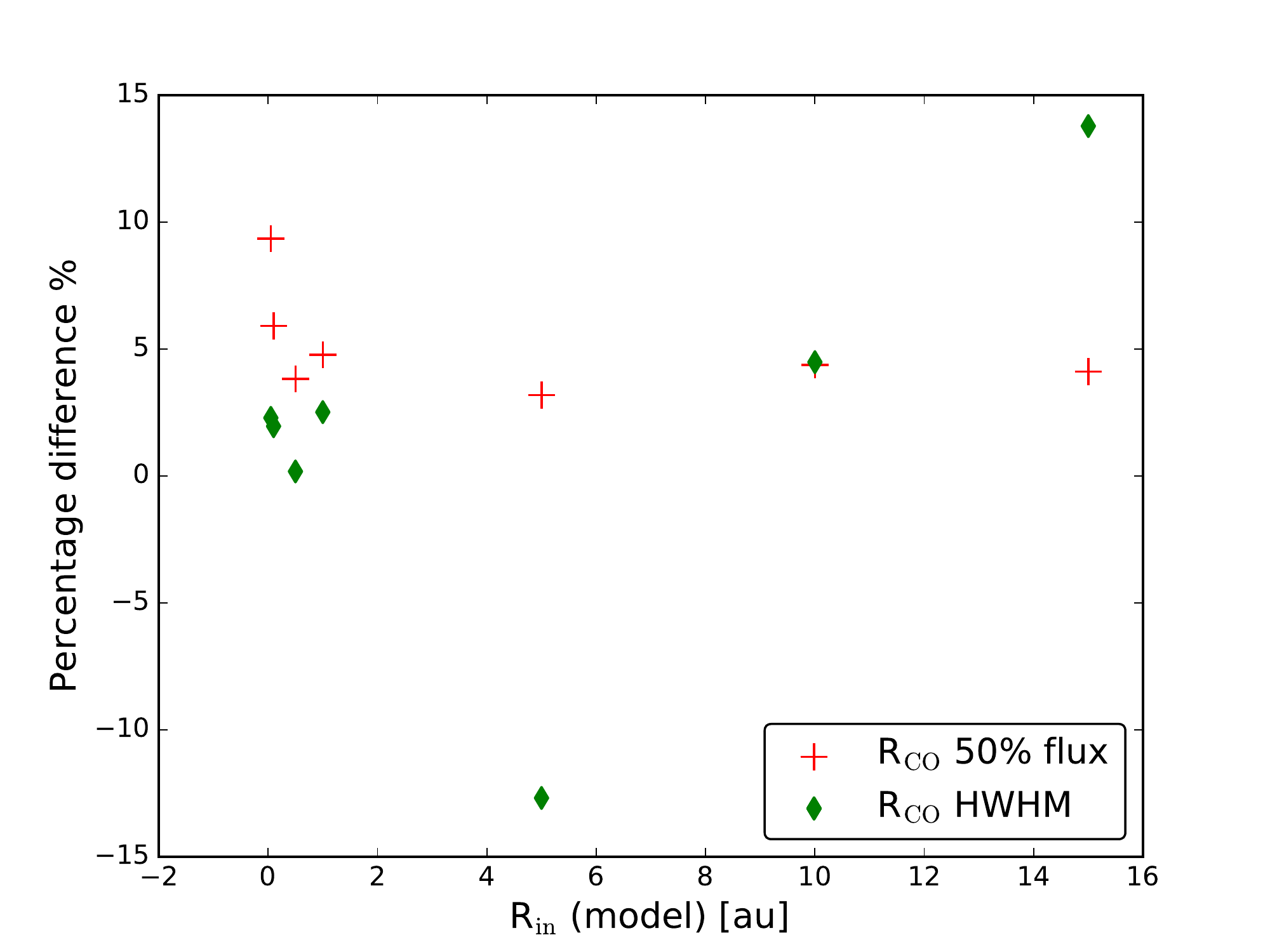} 
\caption{
Plots of the difference between radius definitions and inner radius $\left( (R_\mathrm{X}-R_\mathrm{in}) / R_\mathrm{in} \right) \cdot 100$. The red crosses are the radius that encloses 50\% of the radially integrated line flux in the models. Green diamonds are taken from the line half width at half maximum as in \citet{bp15}.} 
\label{fig:radius_def}
\end{figure}

Given the complex excitation processes for CO, we also ran two additional $R_\mathrm{in}$ model series including 40 rotational for each vibrational level and 7 vibrational levels for the ground electronic state ($X^1\Sigma^+$), hereafter ''Large CO molecule''. In addition, since CO gets electronically excited from FUV radiation, we ran the same models, including the first electronic state ($A^1\Pi^+$) also with 7 
vibrational levels each consisting of 40 rotational levels (''Large CO molecule with fluorescence'').
The last series thus includes fluorescence as additional pumping mechanism for the excited ro-vibrational levels of CO \citep{thi13}. All models, including the ones with ''small CO molecule'', consider automatically the IR pumping from dust thermal emission and collisional quenching.
We calculated a detailed line radiative transfer only for a sub-selection of CO transitions with these larger CO molecules (Table~\ref{Enea}). These synthetic line profiles were also used to test our method of extracting the emitting radius for comparison to the observations, as explained in Section~\ref{2.3}.

\subsection{Model parameters extracted for comparison to the data}
\label{2.3}

From the models, we extract properties of the emission that can be compared to those extracted from the observations. For the observations, \citet{bp15} measured the CO vibrational ratio and emitting radius from stacked lines that 
are free from blending with other lines, specifically lines around P(10) for the $v_{1-0}$ lines, and lines around P(4) for the $v_{2-1}$ lines. From the models, we therefore take the vibrational ratio between the P(4) $v_{2-1}$ and P(10) $v_{1-0}$ model lines. The emitting radius ($R_\mathrm{CO}$) is defined as the distance from the central star where the radially integrated model line flux reaches 50\% of the total CO~$v_{2-1}$ P(4) $v_{2-1}$ or P(10) $v_{1-0}$ line flux, whichever line is the less radially extended. 
In order to be consistent in the comparison with the observational analysis by \citet{bp15}, in which the narrow line component, where present, has been separated from the broad component tracing the innermost disk region, we also reduce the radial integration of the line flux for the line emitted from the most radially extended region. This is done integrating the cumulative radial line flux according to the line with less extended region (either P(4) $v_{2-1}$ or P(10) $v_{1-0}$).
We use this simplified approach, since it is very computationally expensive to produce detailed line profiles for these lines for the entire grid of models. 
We tested that the CO radii derived from the radially integrated flux are similar to those derived from the line velocity at HWHM (Half Width Half Maximum) as in the observations. Figure~\ref{fig:radius_def} shows the model series (4), using the line profiles from detailed line RT as explained in Section~\ref{2.2}, and illustrates that the maximum deviation between the two definitions of $R_\mathrm{CO}$ is at most 15\%, and typically less than 5\%.

\begin{figure*}
\includegraphics[width=0.5\textwidth]{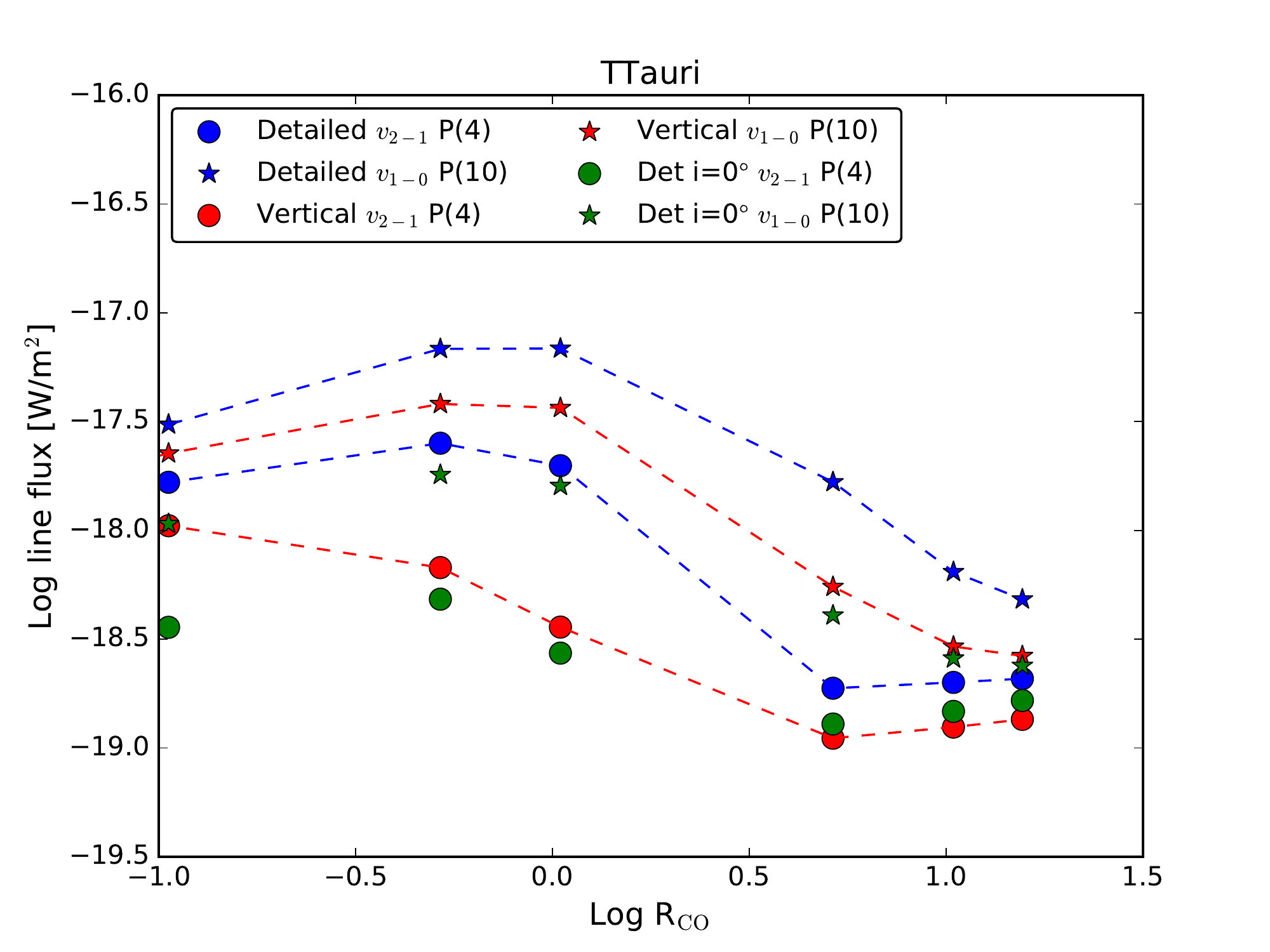}
\includegraphics[width=0.5\textwidth]{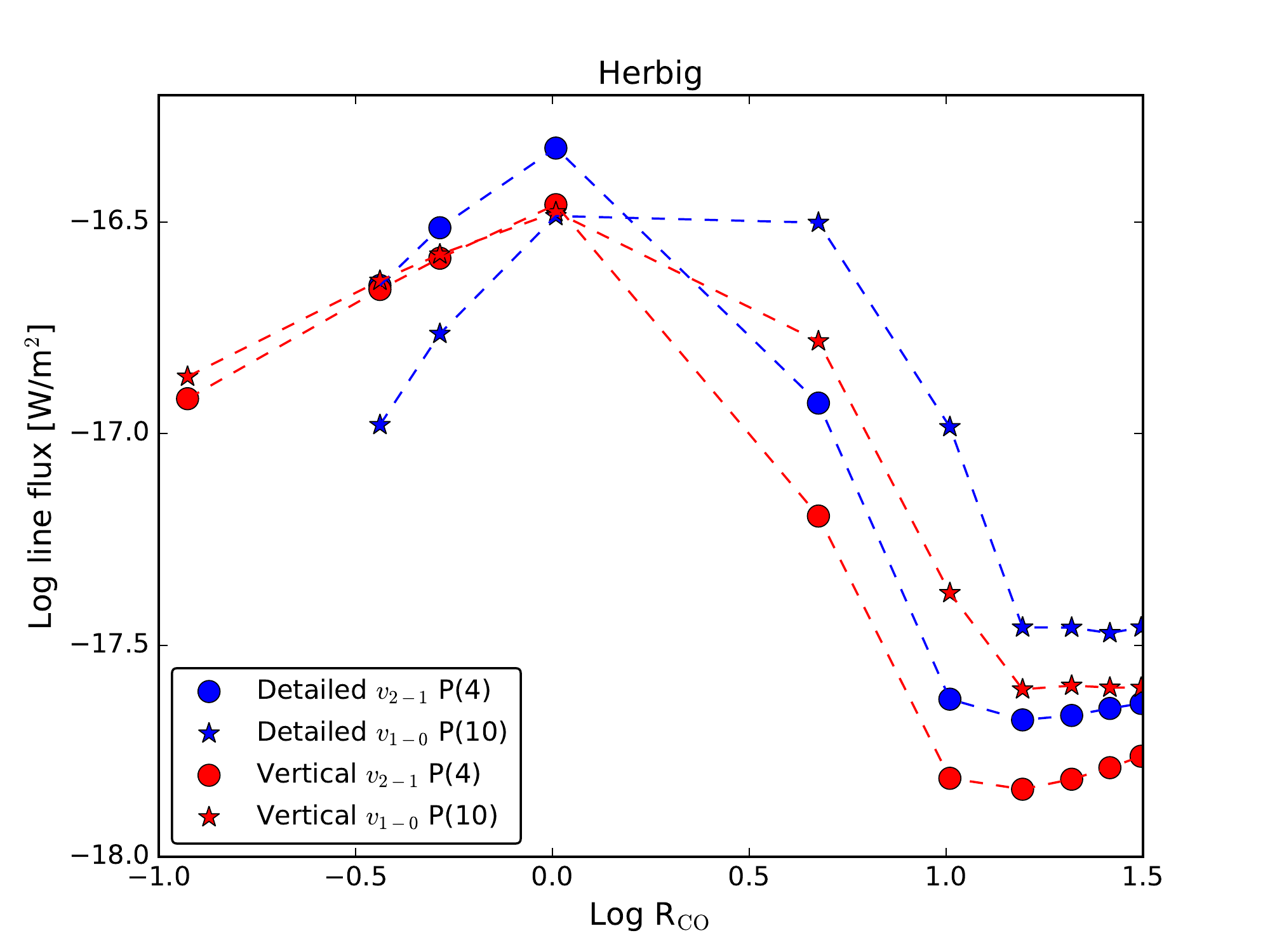} 
\caption{Line fluxes for the $v_{2-1}$ P(4) transition and the $v_{1-0}$ P(10) transition, comparing vertical escape flux (star symbols) and detailed line radiative transfer (circles) for the $R_{in}$ model series. The left plot is for 
the TTauri disk case with fluorescence. The right plot is for the Herbig disk case with fluorescence. The left panel shows the effect of the inclination from the detailed line radiative transfer also for
a face on disk (i~=~0$\degr$). } 
\label{VS}
\end{figure*}

\subsection{Vertical escape line flux versus detailed radiative transfer and disk inclination}

The results presented from our modelling are based on vertical escape line fluxes. This approach is conceptually equivalent to a line flux computation from a face-on disk, and it does not take into account the effect of the inclination. For our subset of models with an extended number of CO levels and fluorescence, we computed detailed radiative transfer for the two ro-vibrational lines of our study (Fig.~\ref{VS}). The fluxes produced by this computation are in agreement with the vertical escape fluxes within about 0.5~dex. Major deviations occur only in the models with intermediate sized gaps. The trend followed by the different lines is very similar, and thus the line ratios will suffer marginally from the detailed effects of the inclination. This makes our approach robust and independent from disk inclination. Furthermore, considering the line ratios instead of the individual flux provides a proxy for the vibrational temperature of the gas, and given the close wavelengths of the two transitions, it makes this value independent from the continuum flux in the near-IR. In the line emitting regions of the two lines considered here, the gas density is much lower than the critical density of the line and hence these two ro-vibrational transitions are in non-LTE (Fig~\ref{Tn}). The need for a non-LTE treatment was already previously found by e.g.\ \citet{Brittain2009}, \citet{thi13}, \citet{bertelsen}, and \citet{bosman}. In agreement with these studies, we find that the density in the emitting region is getting farther from the critical density with increasing $R_\mathrm{in}$ and so $R_\mathrm{CO}$. \citet{bosman} claim that this is an essential condition to get a high 
ro-vibrational ratio in models with larger cavities.

Line fluxes produced from our models are not meant to match the observed ones, as the IR continuum is not fitted to actual data of the observed sample
Therefore, our models should not be considered for line flux predictions. The Herbig disk models do show line fluxes that agree with those typically observed in Herbig disks \citep[Table~6 of][]{banz17}. Instead, our TTauri models produce line fluxes that are $\sim$1~dex weaker than those observed. Below, we explain some of the factors that could reduce this discrepancy.

The plots with detailed line fluxes are from models with the fluorescence pumping mechanism included, but we verified that there is not a significant effect from adding fluorescence or considering a larger set of energy levels on the line ratios, compared to the small standard one used in the majority of our models (Table~\ref{Enea}).

\begin{figure*}
\includegraphics[width=0.5\textwidth]{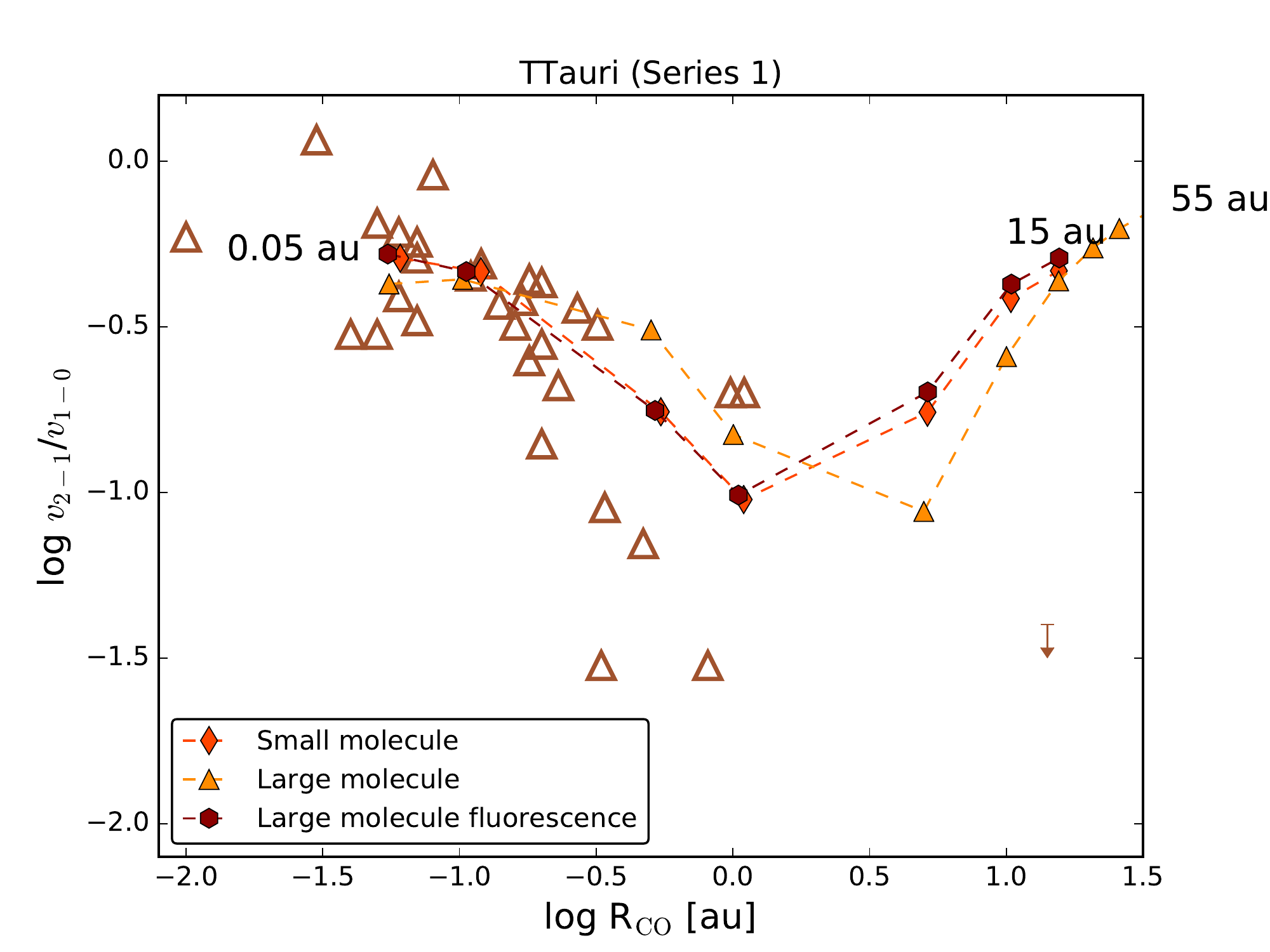}
\includegraphics[width=0.5\textwidth]{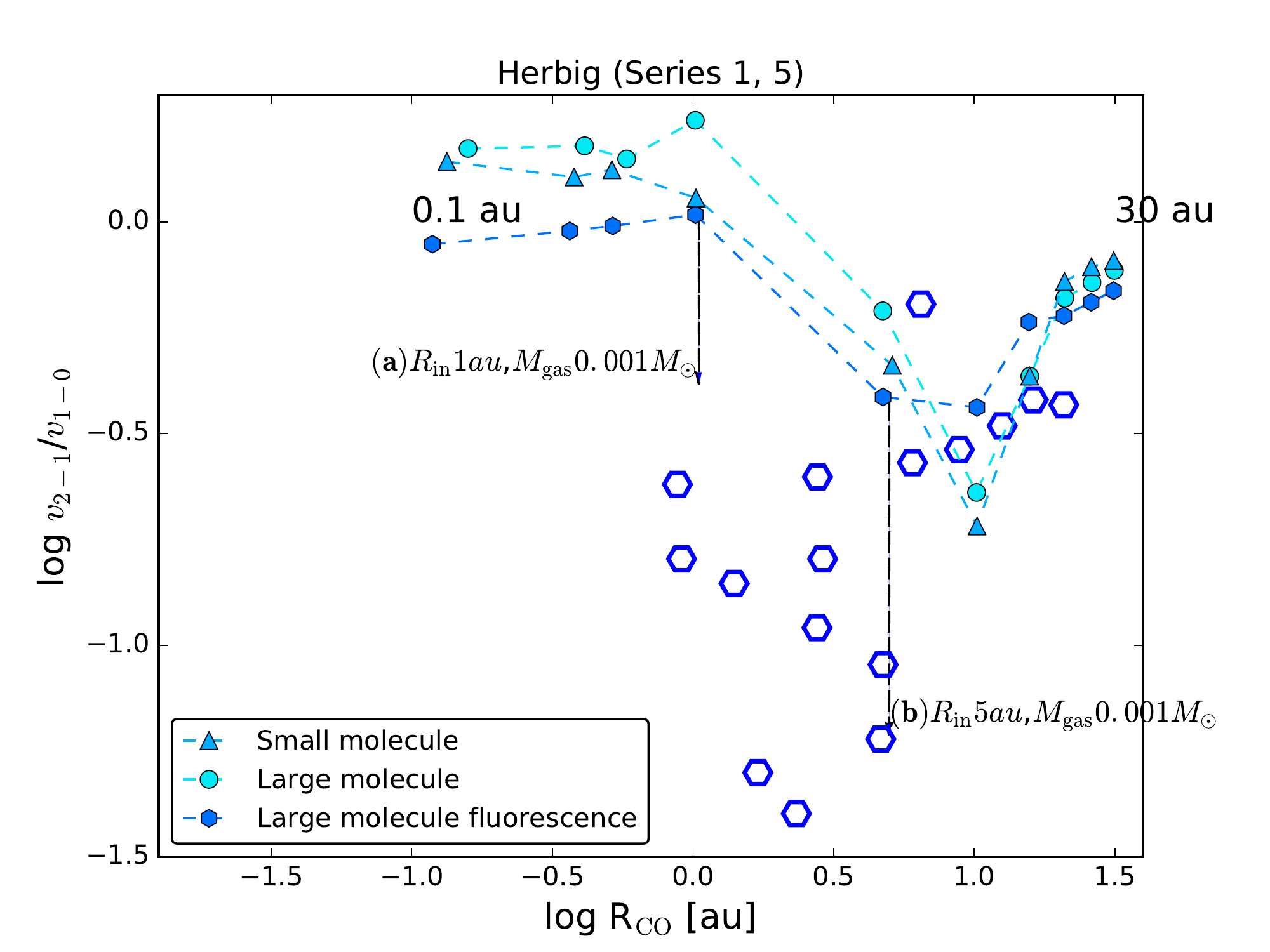} 
\includegraphics[width=0.5\textwidth]{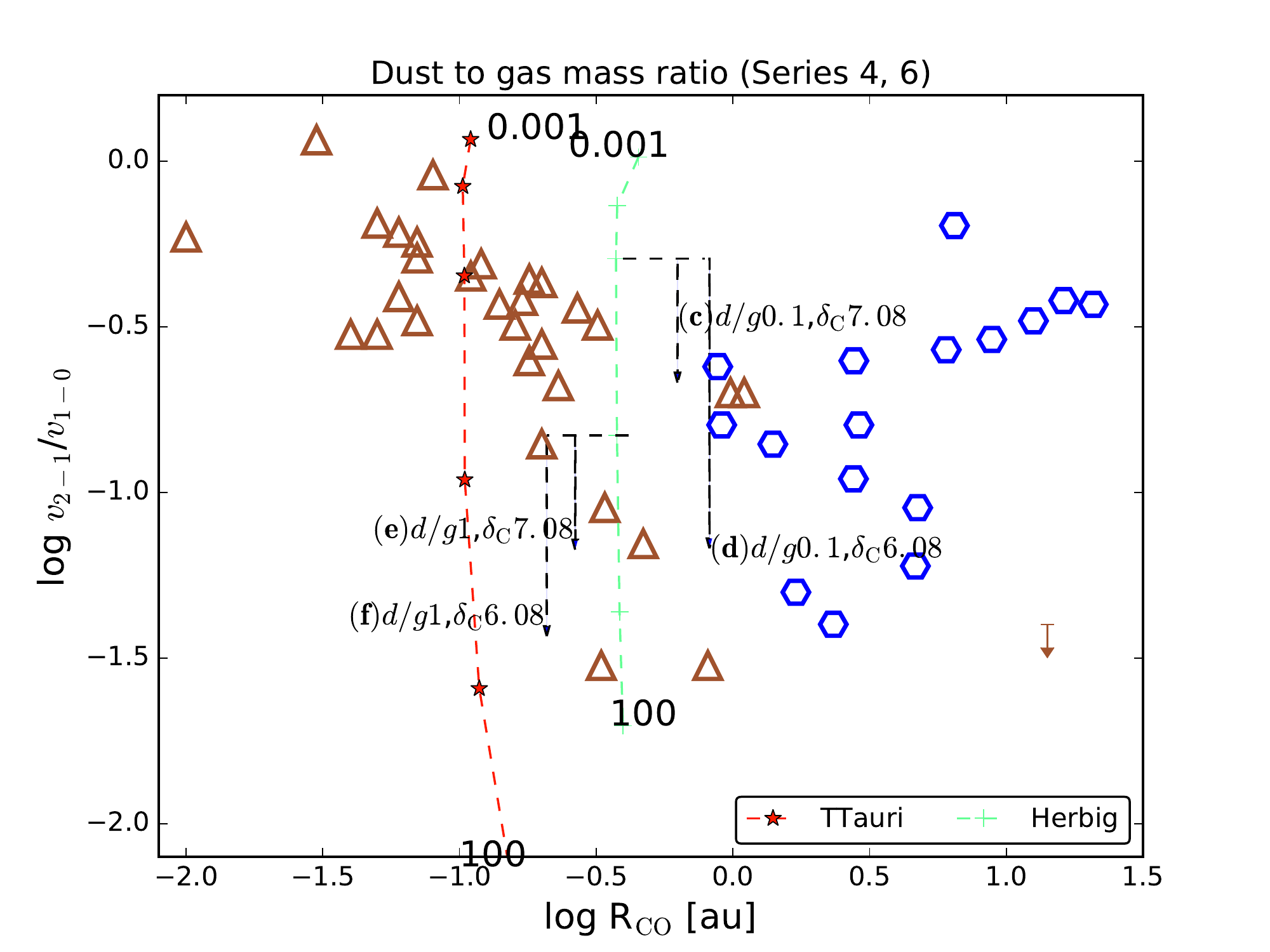} 
\includegraphics[width=0.5\textwidth]{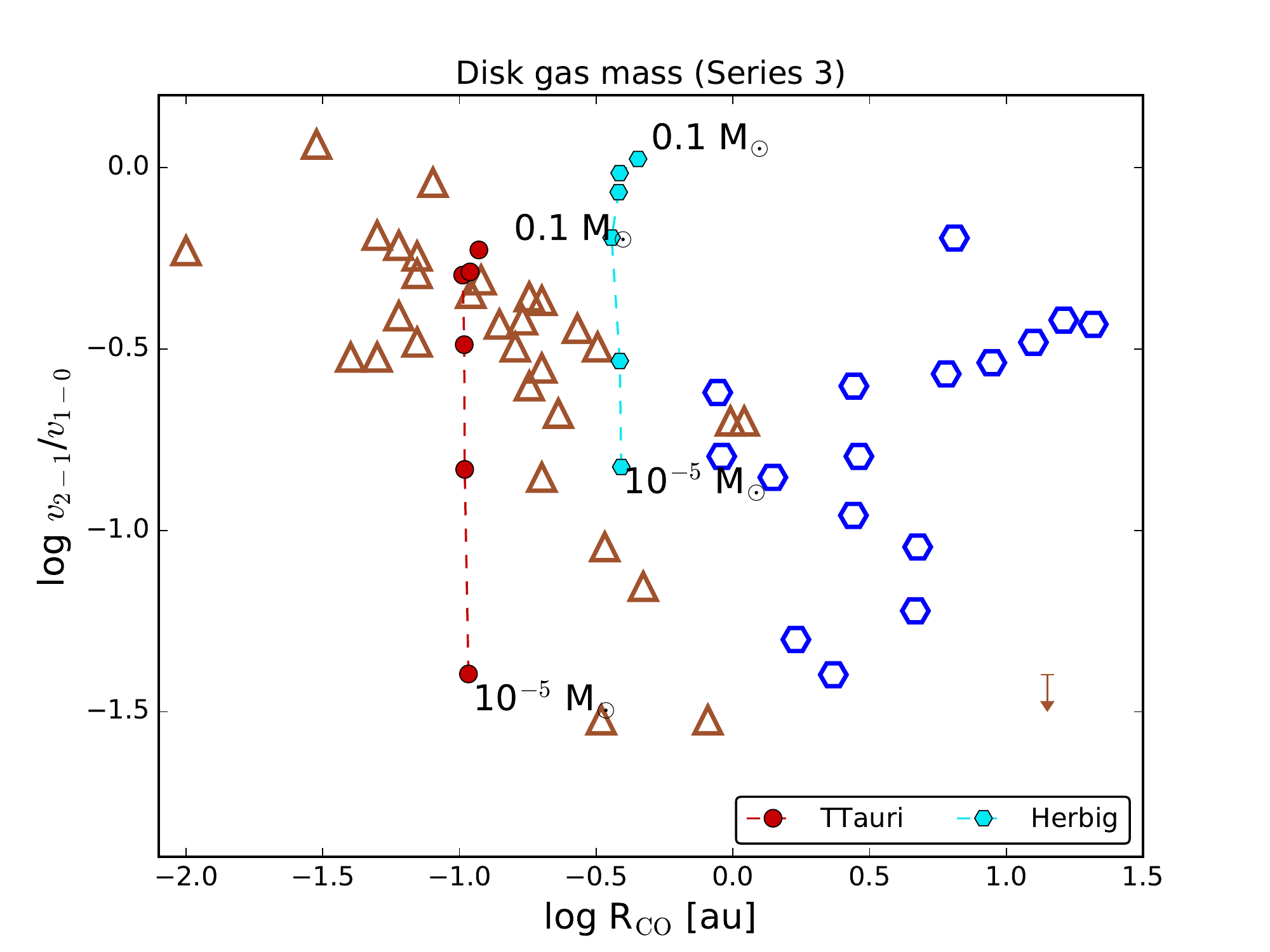}
\includegraphics[width=0.5\textwidth]{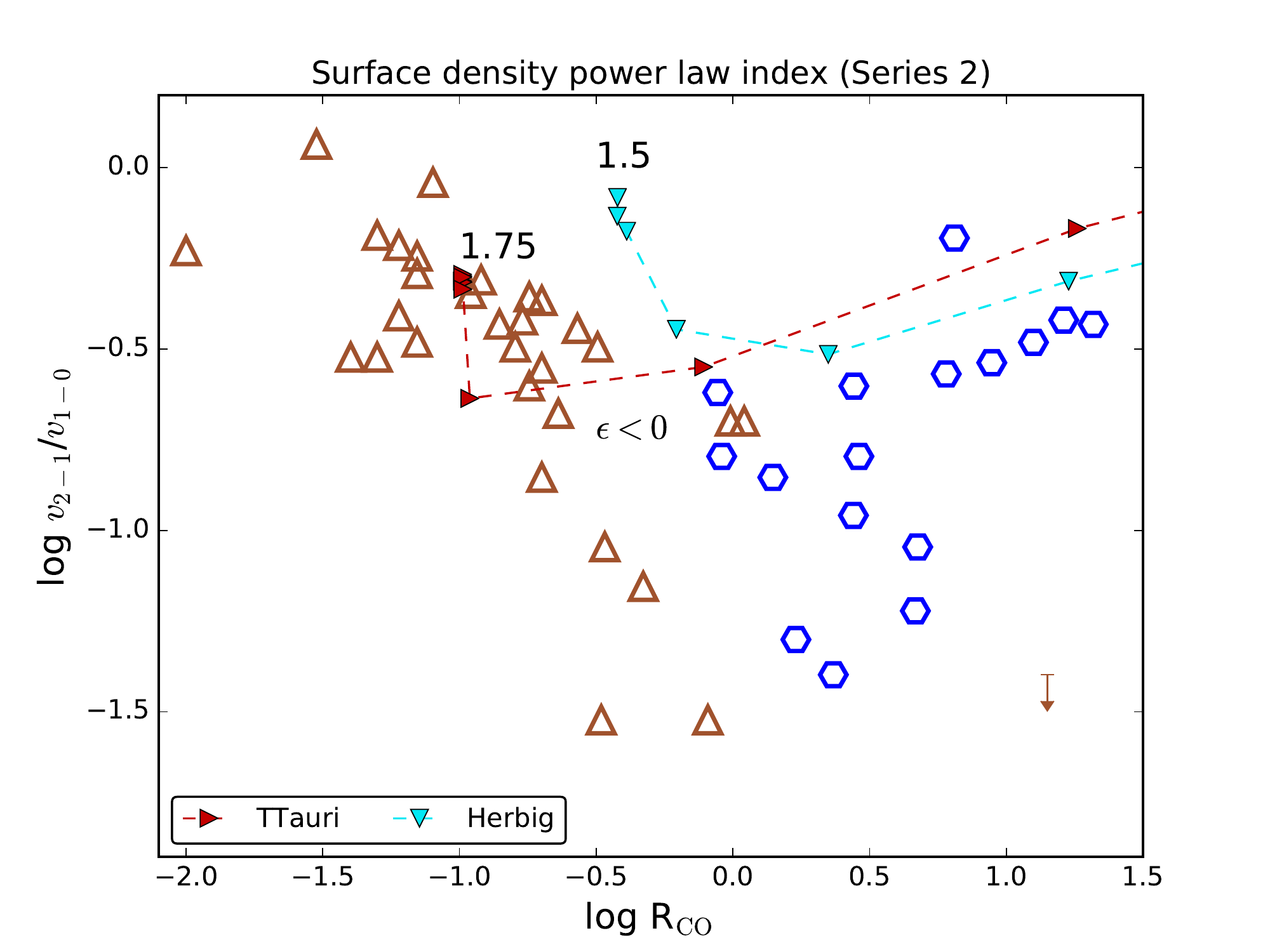} 
\includegraphics[width=0.5\textwidth]{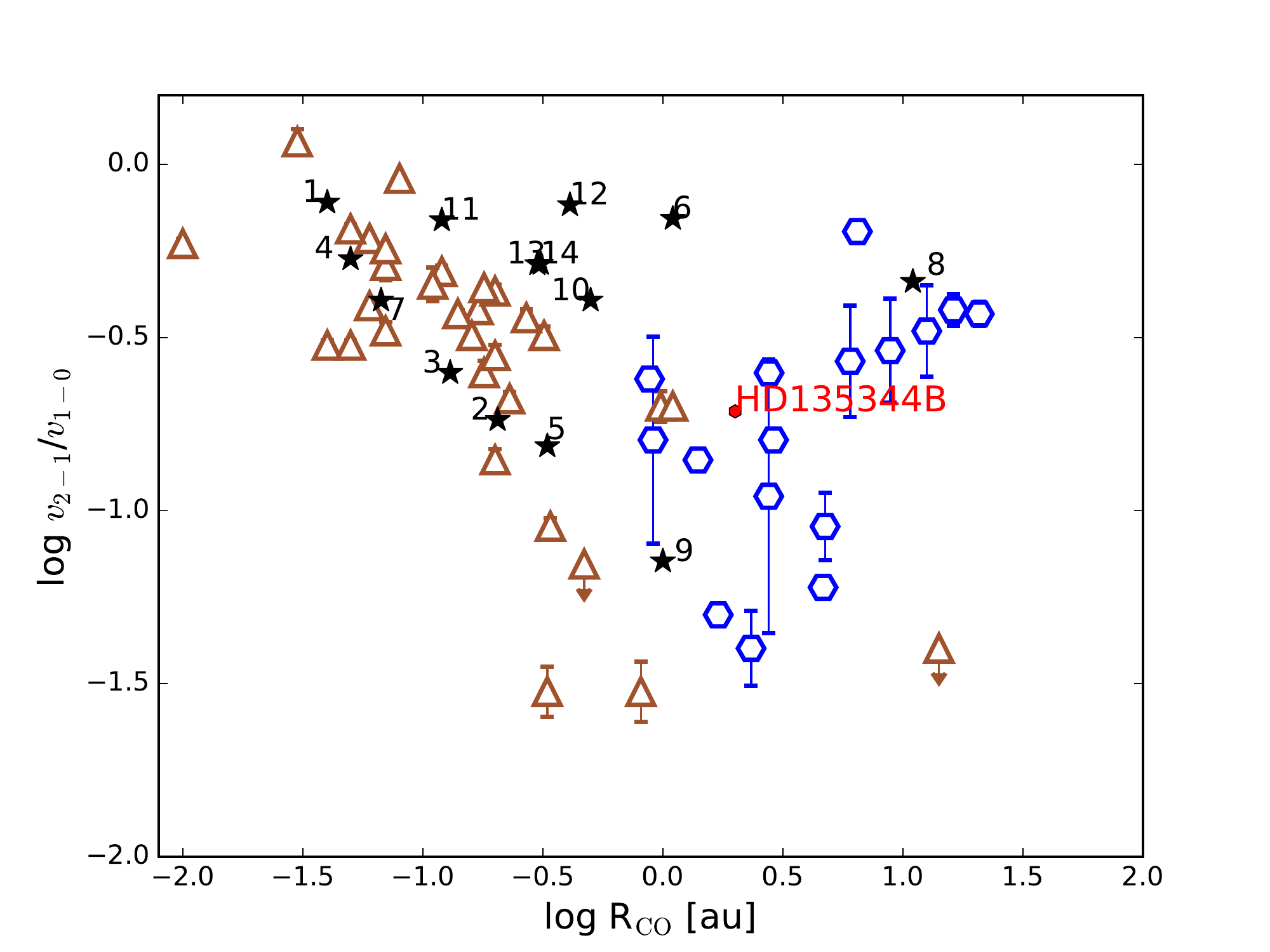}
\caption{Vibrational ratios $v_{2-1}$/$v_{1-0}$ versus $R_\mathrm{CO}$, comparing observations and models. 
Observations are separated into TTauri (brown open triangles) and Herbig (blue open hexagons) disks, and taken from \citet{bp15} and \citet{banz17}. The first two top panels show the TTauris and Herbigs separated (Series 1), all other panels combine them. The top right panel includes also the effect of parameter combinations; the vertical arrows indicate the line ratio produced by models with the same $R_\mathrm{in}$, but different gas mass (Series 5 and 6 in Table~\ref{Overview}). The two panels in 
the middle show how the dust-to-gas mass ratio (Series 4) and the mass of the gas (Series 3) influence the line ratios. In the middle left panel, the vertical arrows indicate the line ratio produced by models with a given dust-to-gas mass ratio (labelled), but different elemental fraction of carbon (Series 5 and 6, Table~\ref{Overview}). The left bottom panel shows the effect of a different surface density power law index (Series 2). The bottom right panel shows DIANA models of individual disks compared to the observational data (including error bars for the Herbig disks). Individual disk models (black stars) are numbered according to Table~\ref{DIANA}. The red hexagon symbol shows the model of \citet{carmona}}.
\label{fig:model_series}
\end{figure*}

\section{Comparison with observations} \label{sec:res}

In Figure~\ref{fig:model_series}, we show our model series from Table~\ref{Overview} to illustrate how the individual model parameters affect the observed CO emission properties.
The inner disk radius and the surface density power law index (series 1 and 2), are the only two parameters that we explore here able to change both the line ratio and the $R_\mathrm{CO}$.
We find that a variation in gas mass (series 3) produces a change of the CO ro-vibrational ratio (the ratio is larger for models with larger $M_\mathrm{gas}$), while the emitting radius $R_\mathrm{CO}$ is almost completely unaffected. Similarly, a variation in dust-to-gas mass ratio (series 4), affects almost exclusively the line ratio (larger d/g, lower line ratio). The same effect of varying the dust-to-gas mass ratio and $M_\mathrm{gas}$ can be mimicked by changing the elemental abundance of carbon (Fig.~\ref{fig:model_series}, bottom right panel). The sensitivity of the ro-vibrational ratio to parameters related to disk structure and composition, suggests that disks around TTauri stars could have different properties than expected (e.g.\ lower d/g, lower $\delta_\mathrm{C}$, higher $M_\mathrm{gas}$

The models with increasing $R_{in}$ also show that the ro-vibrational ratio decreases with the CO emitting radius until $\sim$~1~au. Beyond this radius, the ratio increases, thus reversing the trend. This upturn is also seen in the observations when TTauris and Herbigs are considered together (top left panel of Fig.~\ref{Overview}). In the models, both line fluxes of P(4) $v_{2-1}$ and P(10) $v_{1-0}$ first increase and then drop with increasing the size of the cavity, but the rate at which it happens is different (see Figure~\ref{VS}).

Our model explorations provide theoretical support to the interpretation in \citet{bp15}, that disks with larger $R_\mathrm{CO}$ can be explained by ongoing processes of disk evolution that deplete inner disks from gas, such as inside-out "carving" in both in gas and dust, or more complex partial depletion of the two components (i.e.\ a lower $M_\mathrm{gas}$ and/or lower dust mass, coupled to a small inner disk cavity). 
Our model series for the disks around TTauri stars agrees well with the observations at $R_\mathrm{CO}\!<\!1$~au (Fig.~\ref{fig:model_series}), showing that the ro-vibrational line ratio should decrease by about 1.5 dex with increasing cavity size.
The model series predict that at larger radii the line ratio should increase again. 
The vertical spread of the data points can be explained from the combined effect of other parameters, such as dust-to-gas mass ratio and disk gas mass (Fig.~\ref{fig:model_series} middle right and left). 


When we consider Herbig star disks, the situation becomes more complicated, and our model series with an inner gap (series 1) cannot reproduce the observations of targets with low ro-vibrational ratios. 
The significant mismatch in the line ratio predictions of our model series 4 can be as large as $\simeq\!1$~dex (top right panel of Fig.~\ref{fig:model_series}). This could be improved considering the combined effect of parameters as discussed previously, by including the effects of a significant dust enhancement or strong gas depletion (up to 1-2~dex; magenta stars in the bottom right panel of Fig.~\ref{fig:model_series}). This is a clue that the observed inner disks around Herbig stars may be different from those in TTauri stars. The inner zones are potentially not just empty, but more likely partially depleted in gas or show an enhanced dust content (keeping the same gas mass). This could then contribute to producing a lower CO ro-vibrational ratio.
A very similar vertical displacement is obtained by lowering the carbon abundance ($\delta_\mathrm{C}$) by one dex; we also considered a change in gas mass combined with this carbon depletion (yellow squares and red stars in the bottom right panel of Fig.~\ref{fig:model_series}). 
Additional Herbig disk models varying a combination of parameters, show that such combinations would produce an even stronger variation of the ro-vibrational ratio, than expected from a linear combination of the considered individual parameters. Therefore, a model with $R_\mathrm{in}\!=\!5$~au has a ro-vibrational ratio that is almost one dex lower than the correspondent model with standard $M_\mathrm{gas}$, while for a model truncated at 1~au, the vertical shift increases less than a factor three (Figure~\ref{fig:model_series}, bottom right panel).\newline
\indent
The models used here have a monolithic disk structure where the properties (e.g. the g/d ratio) are varied for the whole disk. This is clearly just an approximation, because Herbig disks often show gaps and ring structures, i.e. a radial dependence of dust and gas distributions.
 However, the CO lines we focus on are emitted in the inner disk, so that the outer disk structure does not matter for this study. In the inner region, conditions can become quite extreme as a result of disk evolution, justifying e.g.\ very high values for the dust-to-gas mass ratio, i.e. very low gas mass with a 'primordial' value for the disk dust mass, or viceversa disk regions entirely depleted in dust. Some evidences for such extreme disk conditions with dust devoid regions and/or regions with dust content equal to the gas content have been found from 
 the interpretation of near-IR and submm observations using disk models
 \citep[e.g.][]{carmona,carmona1,vandermarel3}. 

\section{Discussion}\vspace{5mm}

\subsection{Exploring more complex inner disk structures in Herbigs}\vspace{5mm}

In Fig.~\ref{fig:model_series} bottom right, we included the line ratio extracted from disk models for the objects reported in Table~\ref{DIANA}, which were fitted following the DIANA\footnote{https://dianaproject.wp.st-andrews.ac.uk} standard \citep{woitke4}. These objects are a heterogeneous collections of 14 disks \citep[both TTauris and Herbigs,][]{woitke6}, which match the multi-wavelength continuum and the available line fluxes, plus additional data when available \citep{dionatos2019}. 

\begin{table*}
\caption{DIANA standard fit targets}
\centering
\begin{tabular}{l|c|l|l|l|l}\hline\hline
Name & no. & central star & disk type & Notes & fit level\\ 
 &  & type & & & \\ \hline
CY~Tau & 1 & TTauri & disk hole (3.3~au) & puffed-up inner disk with higher z/r & SED+Lines$^*$ \\
GM~Aur & 2 & TTauri & disk hole (15~au) & - & SED+Lines$^*$ \\
LkCa~15 & 3 & TTauri & disk hole (30~au) & inner disk with higher z/r & SED+Lines$^*$ \\
USco~J1604-2130 & 4 & TTauri & disk hole ($\sim$~50~au) & huge cavity & SED+Lines$^*$ \\
HD~97048 & 5 & Herbig & disk hole ($\sim$~50~au) & huge cavity & SED+Lines$^*$ \\
HD~169142 & 6 & Herbig & disk hole ($\sim$~15~au) & puffed-up inner disk & SED+Lines$^*$ \\ \hline
BP~Tau & 7 & TTauri & (pre-)transitional & inner disk with higher z/r & SED+Lines$^*$ \\
DM~Tau & 8 & TTauri & (pre-)transitional & inverted surface density profile & SED+Lines$^*$ \\ 
& & & & in inner disk; inner disk with higher z/r\\  
TW~Hya & 9 & TTauri & (pre-)transitional disk & inverted surface density profile in inner disk & SED+Lines$^*$ \\
AB~Aur & 10 & Herbig & (pre-)transitional & - & SED+Lines$^*$ \\
HD~142666 & 11 & Herbig & (pre-)transitional & inner disk with higher z/r & SED+Lines$^*$ \\
HD~163296 & 12 & Herbig & (pre-)transitional & puffed-up inner disk with higher z/r & SED+Lines$^*$ \\ 
& & & & than the inner disk & \\\hline
MWC~480 & 13 & Herbig & monolithic disk & - & SED+Lines$^*$ \\
RECX~15 & 14 & TTauri & monolithic disk & $R_\mathrm{out}~=~7.53~$au & SED+Lines$^*$ \\ 
\hline
\end{tabular}
\tablefoot{$^*$ Gas line fitting has been done without including CO ro-vibrational lines.}
\label{DIANA}
\end{table*}

\noindent
In DIANA, disks are modeled with more complex inner structures than just a monolithic disk (with the exception of MWC~480 and RECX~15). Several of these disks are consistent with a two zones radial structure, that we labelled as (pre-)transitional, and some others have inner cavities (disk holes).
We have four objects with line ratios below 0.3 (GM~Aur, LkCa~15, TW~Hya, HD~97048), values that our Herbig disk models are unable to reach (Section~\ref{sec:res}).
Only one of these objects is a Herbig star disk (HD~97048), and three of them have a large hole (HD~97048, GM~Aur, LkCa~15). However, these models still do not overlap to the data, in terms of R$_\mathrm{CO}$; only TWHya (\#9) is close to the region that we need to populate with Herbig models, but TWHya is a TTauri. The only Herbig model that falls in the region left empty by our monolithic models is HD135344B, from \citet{carmona}.

Two interesting targets, both TTauris, have the peculiarity of having an inverted surface density profile, and one of the two has the lowest CO vibrational ratio in the DIANA sample (TW~Hya). Disks usually are expected to have a surface density which decreases radially according to a power law with exponent $\epsilon$. However, in certain objects modeling has been consistent with the presence of regions in which the surface density increases radially (DIANA models no. 8 and 9, Fig.~20 and 12 of \citet{woitke6}, and HD135344B, red hexagon, from \citet{carmona} in Fig.~\ref{fig:model_series} bottom right).
Following this result, we considered an additional model series in which we changed the surface density radial distribution for both TTauri and Herbig star disk cases (series 2; Fig.~\ref{fig:model_series} bottom left). The line ratio and the emitting region are not significantly affected by any change of the surface
density radial distribution profile until $\epsilon$ becomes negative (an inverted surface density profile). For negative values, the line ratio initially drops just below -0.5 for both the TTauri and Herbig model series, and the emitting region of CO drifts quickly beyond the inner radius. The ro-vibrational ratio is affected by less than 0.5 dex in both TTauri and Herbig star disks, along the explored dynamic range. This effect partially mimics the series 1. Hence, a combination of gas depletion, inverted surface density index, and/or an inner hole, could in principle explain the observations with large $R_\mathrm{CO}$. 

The potential explanations for such an altered disk structure include the perturbation induced by a Jovian planet in the inner disk \citep{tatulli,lubow,varniere}. In two Herbig star disks consistent with models that show inverted surface density profile (HD~135344B, HD~139614), in the region where the surface density power law index is positive, the gas column density is partially depleted, and the dust column density is strongly depleted. This can be a consequence of the presence of a planet opening a small inner gap \citep{carmona,carmona1}. In our modelling, this is equivalent to a combination of inverted surface density profile and low gas content (and even stronger dust depletion); this combination of parameters would produce a low ro-vibrational ratio as shown in the bottom right panel of Fig.~\ref{fig:model_series} 
ALMA observations support the presence of simultaneous dust and gas depletion in a few disks, based on emission from $^{13}$CO and C$^{18}$O \citep{vandermarel2}. 

\subsection{An alternative explanation for the CO sequence upturn}
\label{upturn}

\begin{figure*}
    \centering
    \includegraphics[width=0.49\textwidth]{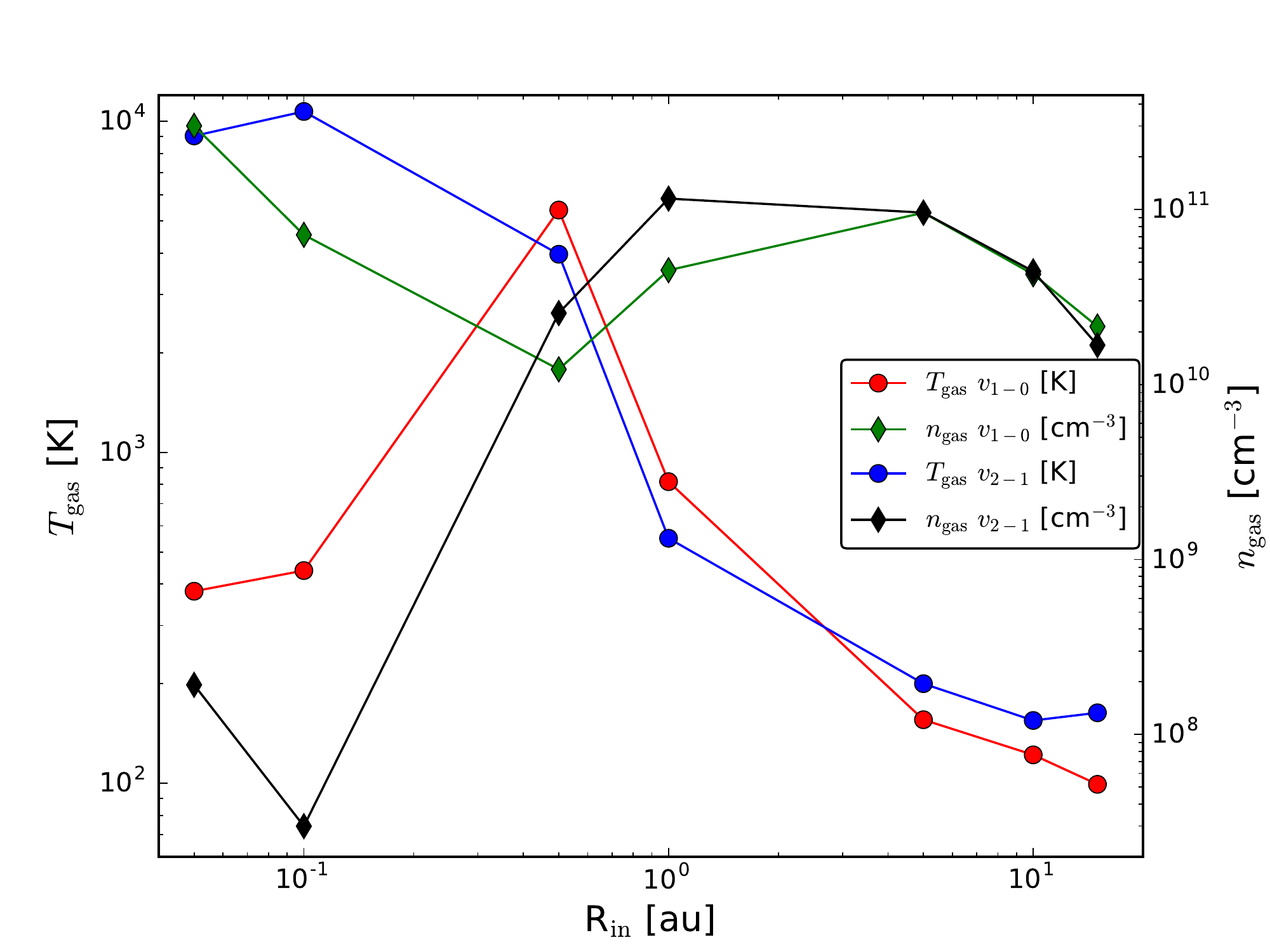}
    \includegraphics[width=0.49\textwidth]{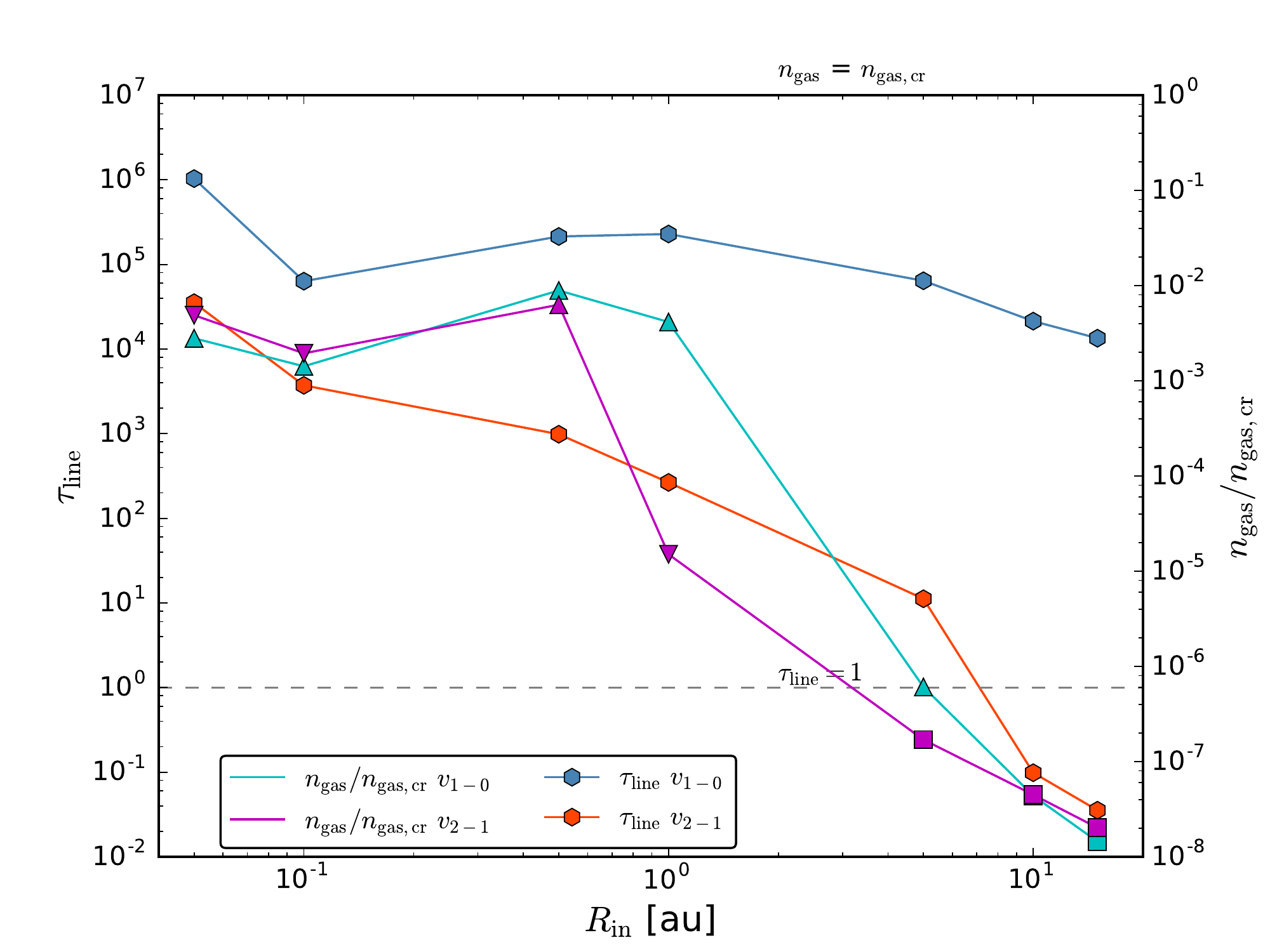}
    \caption{Left: average emitting regions physical conditions $T_\mathrm{gas}$ and $n_\mathrm{gas}$ for the lines P(4) $v_{2-1}$ (respectively red dots and green diamonds) and P(10) $v_{1-0}$ (respectively blue dots and black diamonds). Right: critical density over gas density for the P(10) $v_{1-0}$ (cyan triangles) and P(4) $v_{2-1}$ (magenta triangles). Average line optical depth in the emitting region of the P(10) $v_{1-0}$ (blue hexagons) and P(4) $v_{2-1}$ (orange hexagons) lines. The grey horizontal dashed line marks the optical depth $\tau_\mathrm{line}~=~1$.}
    \label{Tn}
\end{figure*}

Considering a larger number of energy levels including electronically excited levels for the CO molecule provides a more complete and realistic description of the excitation processes in the CO molecule. Our CO lines are computed from detailed non-LTE level populations. This is important, as the average density of the disk region from which these lines are emitted is clearly below the critical density for these transitions (Fig.~\ref{Tn}, the emitting region is here consistently truncated as explained in Sect.~\ref{2.3}). $T_\mathrm{gas}$ is also decreasing with $R_\mathrm{in}$, to almost level off beyond 1~au, in agreement with the results of \citet{bosman}. Therefore, CO levels in our models are populated by 'IR pumping', due to the thermal emission from the dust. As the CO molecule absorbs UV radiation in non dissociative electronic levels, this also affects the population of ro-vibrational states. \citet{thi13} show that this can cause 'UV fluorescence pumping' of the energy levels. 
Previous studies for the Herbig star disk of HD~100546 found fluorescence to increase the line fluxes associated to the ro-vibrational transition $v_{2-1}$ of about a factor two, while the $v_{1-0}$ fluxes are unaffected \citep{bertelsen}. Hot bands from higher vibrational levels are affected up to a dex. For TTauri disks, the UV pumping seems to play a role only for the population of very excited vibrational states, as $v=14$ of the ground electronic state \citep{schindhelm}. In this study we analyze only ro-vibrational lines from the first three vibrational states, $v_{1-0}$ and $v_{2-1}$, of the ground electronic state and the populations appear mostly unaffected by fluorescence in the case of TTauris, and only moderately affected in the case of Herbigs (Fig.~\ref{fig:model_series} top left and right panels). In our models, the $v_{2-1}/v_{1-0}$ ratio is therefore less sensitive to UV fluorescence, while higher vibrational levels clearly show fluorescence pumping in some Herbig disks \citep{brittain2,vanderplas}.   


According to our modeling results, the increase in vibrational ratio found in \citet{bp15} for large $R_{\rm CO}$ values in Herbig disks (see Sect.~\ref{sec:intro}) is not attributed to fluorescence pumping, for both TTauri and Herbig star disks. 
Our models show a strong difference in line optical depth for the $v_{2-1}$ and $v_{1-0}$ transitions. While the P(10) $v_{1-0}$ line is optically thick for all the inner radii, the P(4) $v_{2-1}$ line changes from an optically thick regime in the less truncated models, to an optically thin regime for disk inner radii of $\gtrsim\!1$~au, exactly where the upturn happens in Fig~\ref{fig:model_series}. 
This coincides with a significant drop in line flux of the $v_{1-0}$ line. In comparison, the $v_{2-1}$ line shows a smaller drop in line flux and even levels off for $R_{\rm in}\!\gg\!1$~au. As a consequence the ro-vibrational ratio increases for large inner radii.
The reason for this is that the $v_{2-1}$ in the optically thin regime becomes superthermally excited contrary to the $v_{1-0}$ which is sub-thermally excited. Such a superthermal population for higher vibrational levels of CO was already reported by \cite{thi13}.
The lower volume density of the emitting regions in more truncated disks ($R_{\rm in}\!\gg\!1$~au), combined with a higher critical density of the $v_{2-1}$ leads to stronger non-LTE effects.
These effects compensate also the lower gas temperature and lower emitting gas column density for larger inner radii, leading to an almost constant line flux for large $R_\mathrm{in}$.
A similar process occurs with the variation of the surface density power law index, producing the slight upturn observed in Fig.~\ref{fig:model_series}.

Hence, in our modeling framework, we do not need to invoke effects of dust depletion or dust traps as suggested by \citet{bosman} to reproduce the ro-vibrational upturn. 
Our cavity-series models describe very well the downward trend in ro-vibrational ratios with increasing $R_{\rm CO}$ observed towards TTauri star disks, and they reproduce the upturn in the  ro-vibrational ratio for Herbig disks with large inner radii. However, this cavity-series models fail in matching the very low observed ratios for Herbig star disks with inner radii around 1~au. In this case, we have to invoke a more complex disk structure with e.g.\ a dust-to-gas mass ratio different from the canonical value, or inverted surface density profiles.
In Table~\ref{CentralStars}, we compare our model parameters with the DALI ones. Besides small differences in stellar properties, we find the dust settling prescription as one of the main differences potentially able to affect the CO ro-vibrational ratio result obtained by both models. In our models we used the turbulent settling as described in \citet{dubrulle}, and this affects mainly the outer disk (due to the density dependence). The parametrized approach enforces settling across the entire disk structure, thus strongly affecting the continuum opacity in the upper layers of the inner disk \citep{antonellini2}. The surface density inside 50~au is another significant difference able to affect the ro-vibrational ratios in between the two models. 

\begin{figure}
\label{Em}
\end{figure}

\section{Conclusions}\label{sec:d&c}\vspace{5mm}

In this study, we use mostly existing radiation thermal-chemical models to test the interpretation of trends observed in near-IR observations of CO ro-vibrational lines in TTauri and Herbig star disks \citep{bp15}. We consider a series of models in which only the inner disk radius, surface density power law index, dust-to-gas mass ratio, gas mass, has been changed (respectively, series 1, 2, 3 of Table~\ref{Overview}). These parameters have been selected as promising to investigate the proposed explanations for the observed trends between $v_2/v_1$ and $R_\mathrm{CO}$. Additional models with detailed line radiative transfer and a large number of CO energy levels have been included. We summarize here the investigated scenarios and our findings:\vspace{4mm}

\indent
- \underline{Scenario 1:} Inner dust- and gas-devoid cavities of increasing size are able to match the decrease in CO vibrational excitation with increasing radius as observed in disks around TTauri stars. The same models instead fail to match CO emission in Herbig disks overall, but can match a few disks that have large inner cavities ($> 10$~au).
\newline

\indent
- \underline{Scenario 2:} An inverted surface density profile could also produce the trend for both TTauri and Herbig disks, but our models suggest that additional factors (e.g. a dust or gas depletion) should also be combined to try and reproduce the observed low CO rovibrational ratios. 
\newline

\indent
- \underline{Scenario 3:} Additional factors to be considered to match CO emission in Herbig disks include inner cavities that are only partially depleted from gas or dust. Detailed modeling of individual Herbig disks seems to favour a dust-depleted cavity \citep{carmona,carmona1}. This is also supported by observations of CO isotopologues and dust continuum with ALMA \citep{vandermarel2}.
\newline

 




The models produce an upturn and the increase of vibrational ratio with radius, for radii larger than 1~au, without invoking the effect of UV fluorescence. As seen in Fig.~\ref{VS}, both line fluxes decrease until $\sim$~1~au, in the case of the $v_{2-1}$, the decrease stop and the flux level off. Our models show that the $v_{2-1}$ transition becomes optically thin for models truncated beyond 1~au, simultaneously, the involved energy levels in the transition get super-thermally excited, balancing the effect of the line optical depth. This causes a turning point in the ro-vibrational ratio.

The fact that we can match CO emission in TTauri disks with the approximation of completely gas- and dust-devoid inner cavities suggests that their evolution is consistent with an inside-out disk clearing.
As an alternative possibility, if we look at DIANA models of TW~Hya and DM~Tau, this trend is degenerate with the existence of a population of transition disks which show an inverted surface density radial profile, combined with low gas mass. 

For the Herbig star disks, the possibility of more complicated inner structures is supported by the fact that we cannot reproduce most of the observations by varying a single disk parameter. The models get closer to the observations if we combine multiple parameters such as disk inner radius, dust-to-gas mass ratio, disk gas mass. Alternatively, a combination of an inverted surface density profile with different content of gas and dust in disk could also be considered.
We also showed that reducing the carbon abundance can decrease the ro-vibrational ratio by a similar factor as found for a higher dust-to-gas mass ratio or for a lower $M_\mathrm{gas}$.
In order to model individual disks in detail, it is necessary to include additional observations that can help constraining the inner disk structure (e.g.\ interferometry, gas line profiles), as done in \citet{carmona}.

In this work we have explored the effects on the CO ro-vibrational lines from the inner disk produced by individual and combined model parameters. More work is needed in the future to connect the results of this study with physical scenarios of inner disk evolution (i.e. photoevaporation, planet formation) by folding in additional observational data such as near- and mid-IR interferometry, and optical/near-IR emission line profiles.

\bibliographystyle{v611/aasjournal}
\bibliography{DraftAnd}

\begin{acknowledgement}
Astrophysics at Queen's University Belfast is supported by a grant from the STFC (ST/P000312/1)
\end{acknowledgement}

\appendix

\section{CO model molecule}

Table~\ref{Enea} provides an overview of the energy levels used in the different disk model setups for the CO molecule.

\begin{table*}[bt]
\caption{Overview of CO ro-vibrational levels included in the model}
\centering
\begin{tabular}{cc|c|c|c}
\hline\hline 
Electronic state & Vibrational level & & Rotational levels & \\ \hline 
& & Small molecule & Large molecule & Large molecule with fluorescence\\ \hline
& v = 0 & 50 & 40 & 40 \\
& v = 1 & 50 & 40 & 40 \\
& v = 2 & 50 & 40 & 40 \\
$X^1\Sigma^+$ & v = 3 & 50 & 40 & 40 \\
& v = 4 & 0 & 40 & 40 \\
& v = 5 & 0 & 40 & 40 \\
& v = 6 & 0 & 40 & 40 \\ \hline
& v = 0 & 0 & 0 & 40 \\
& v = 1 & 0 & 0 & 40 \\
& v = 2 & 0 & 0 & 40 \\
$A^1\Pi^+$ & v = 3 & 0 & 0 & 40 \\
& v = 4 & 0 & 0 & 40 \\
& v = 5 & 0 & 0 & 40 \\
& v = 6 & 0 & 0 & 40 \\ \hline
\end{tabular}
\label{Enea}
\end{table*}






\section{Line emitting regions}

\begin{figure*}
\mbox{\includegraphics[width=0.4\textwidth]{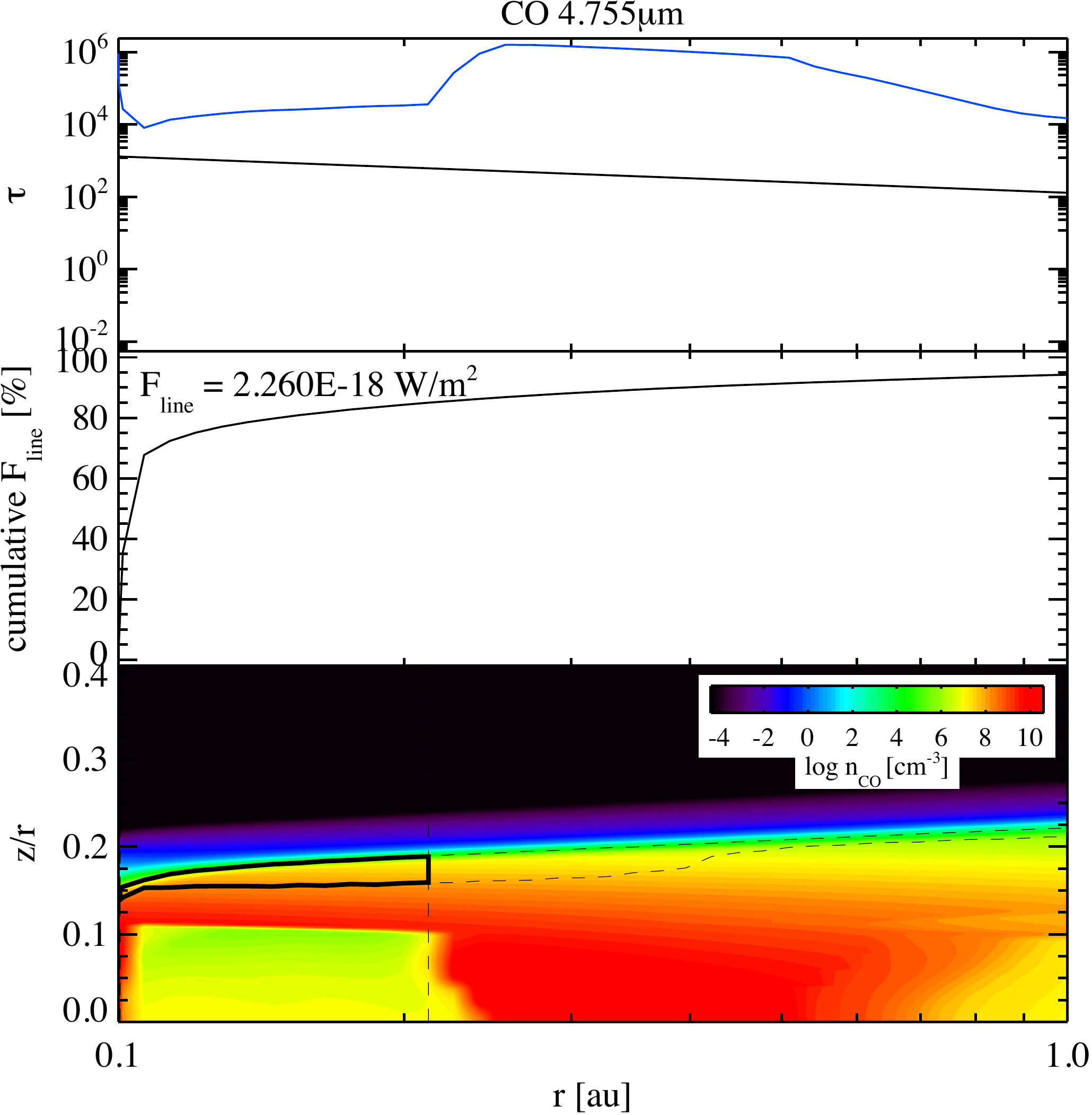} 
\hspace*{3mm}\includegraphics[width=0.41\textwidth]{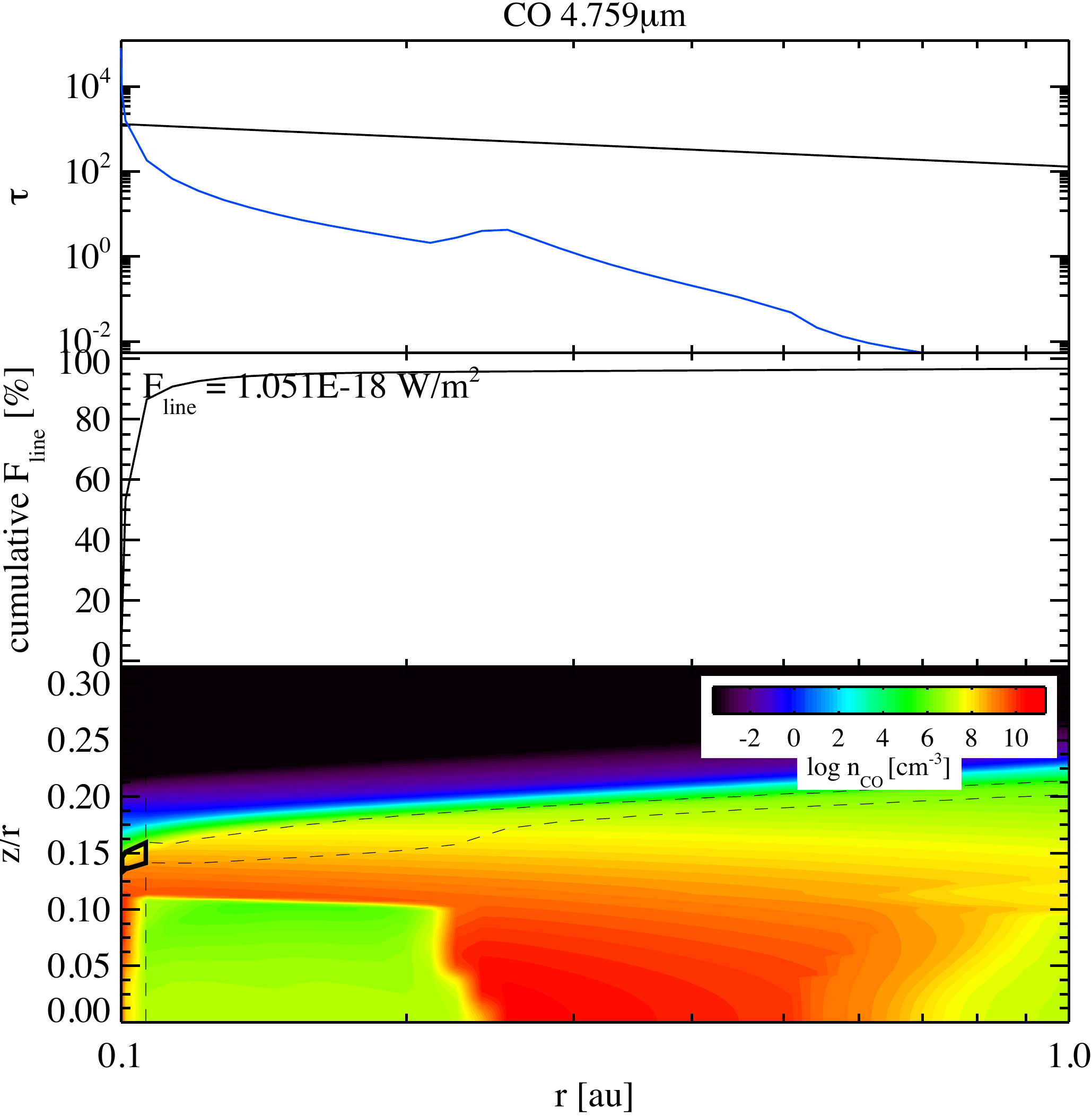}}
\mbox{\includegraphics[width=0.4\textwidth]{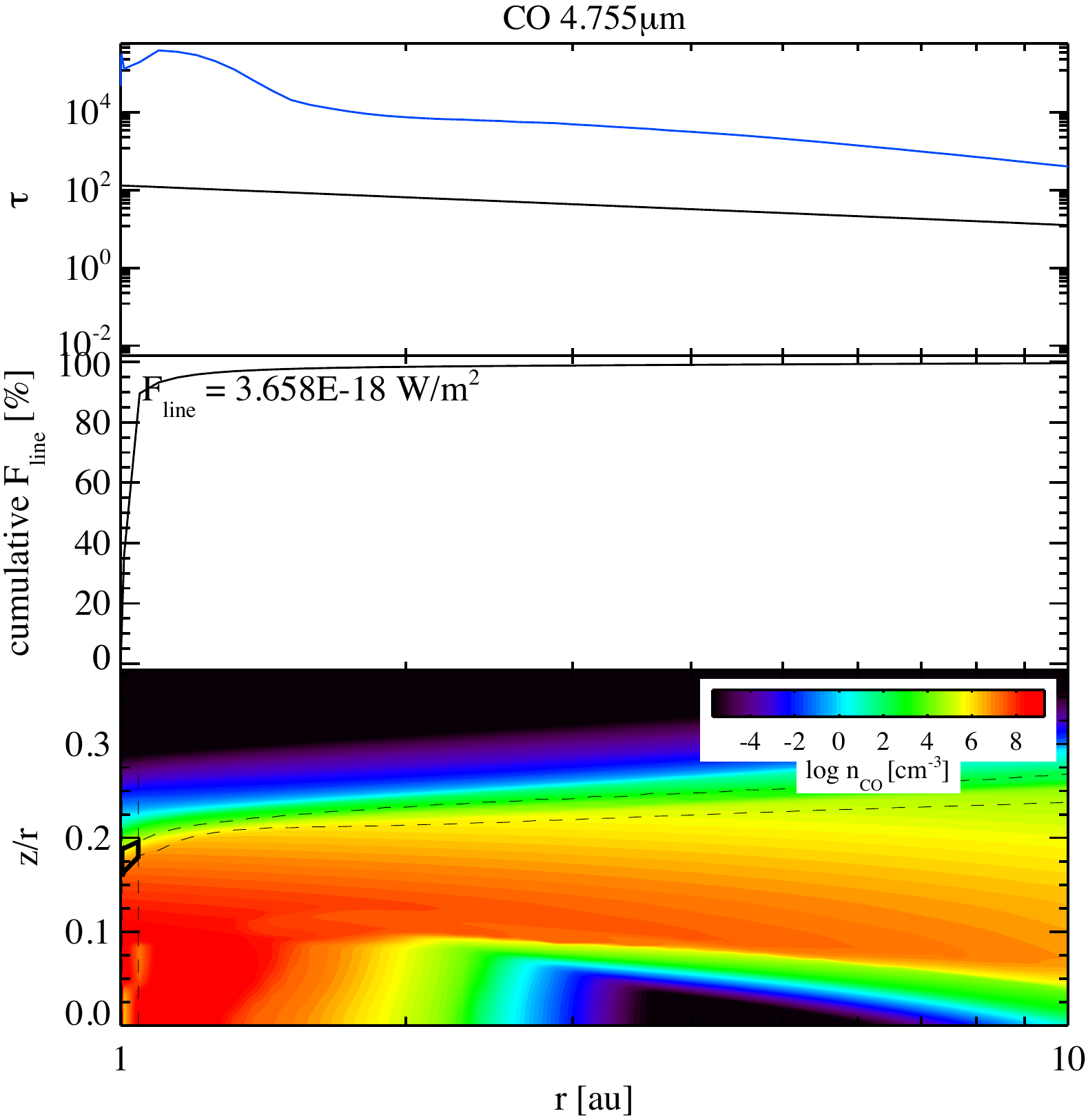} 
\includegraphics[width=0.42\textwidth]{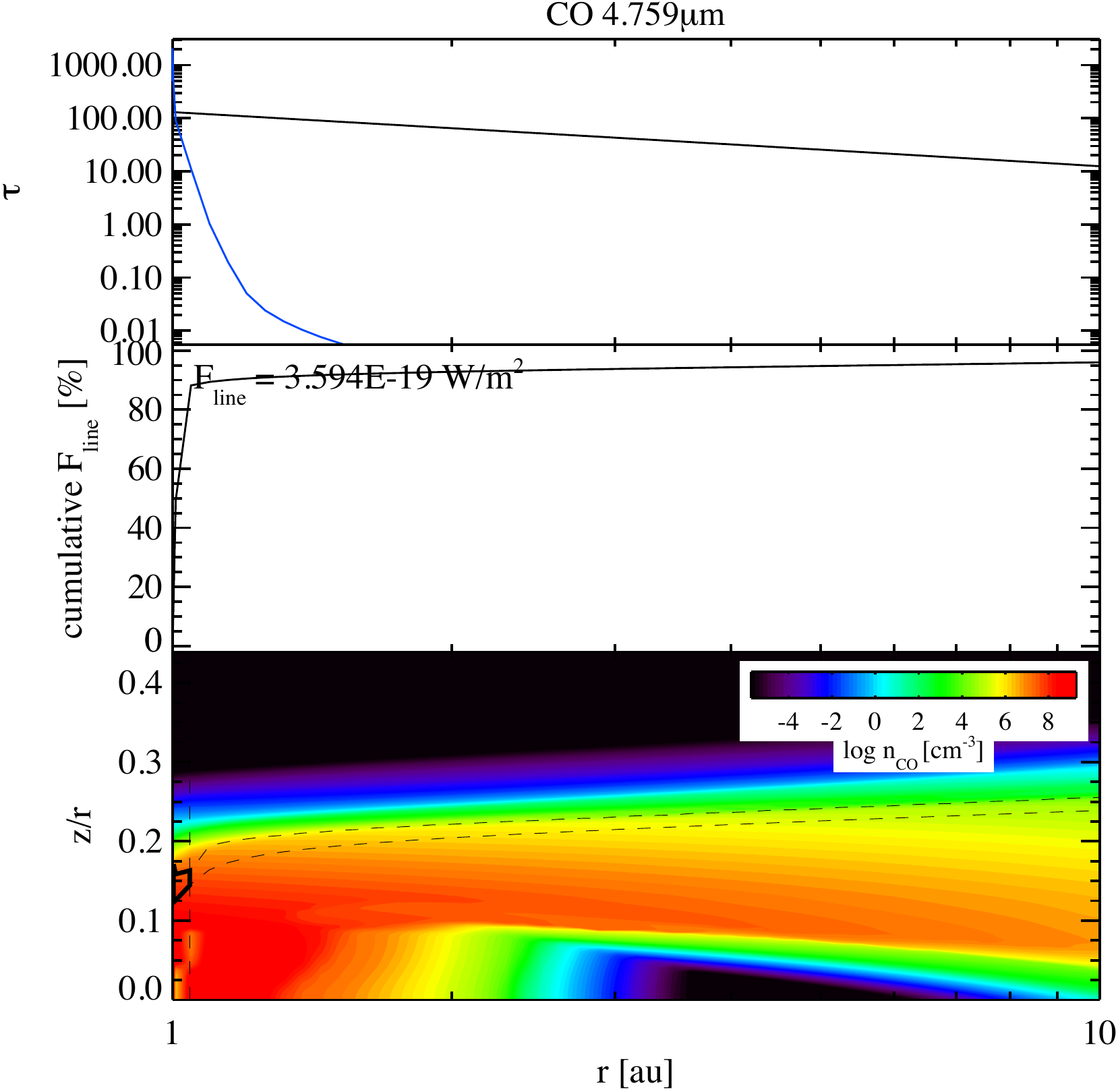}}
\mbox{\includegraphics[width=0.4\textwidth]{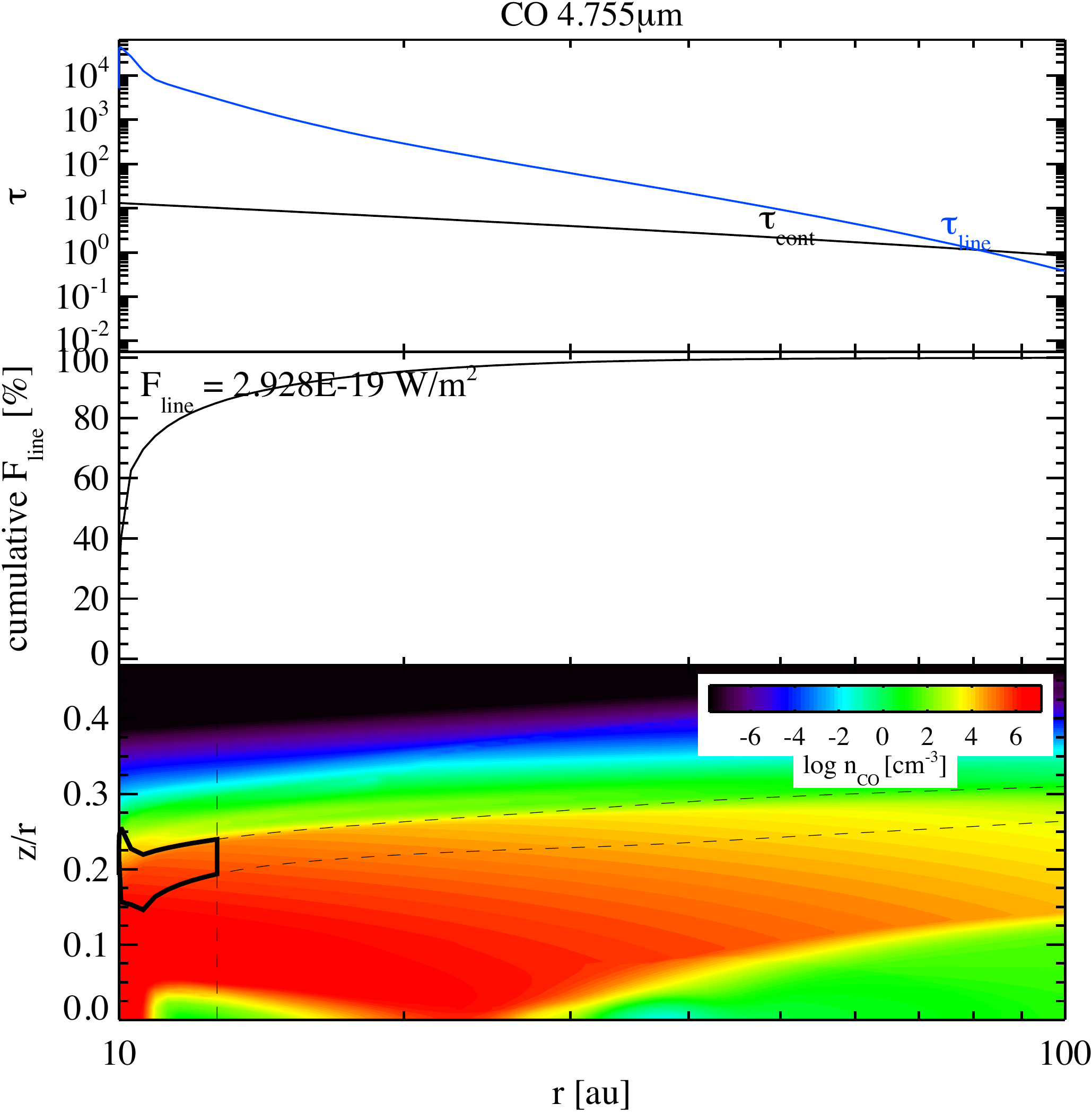}
\hspace*{2mm}\includegraphics[width=0.41\textwidth]{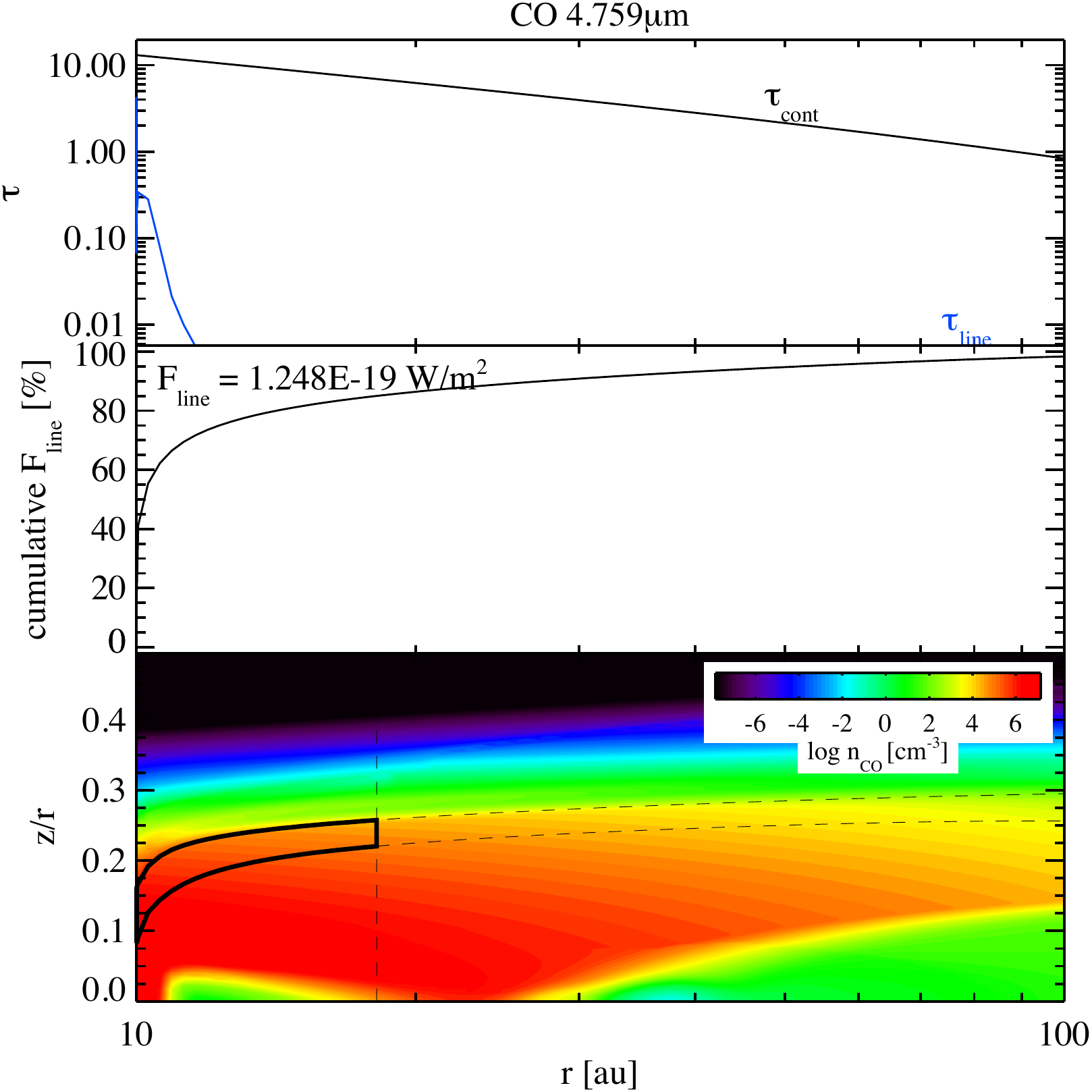}}
\caption{Zoomed-in line emitting regions from the TTauri models with $R_\mathrm{in}\!=\!0.1~$au for the P(10) $v_{1-0}$ (top left) and the P(4) $v_{2-1}$ line (top right). The middle and bottom rows show the same for the models with $R_\mathrm{in}\!=\!1.0~$au and 10~au. In each figure, the top panel shows the continuum optical depth (black line) and the line optical depth (blue line). The middle panel shows the cumulative line flux from vertical escape probability as a function of the radial distance from the star. The bottom panel shows the CO density in color scale with the region from which the vertical * radial integrated flux, amounting then to 49\% of the total line flux are emitted, and reported as numbers in the middle slice of each plot} (black contours, 15-85\% of the vertical and radial integrated line flux, dashed lines). 
\label{Region}
\end{figure*}

Figure~\ref{Region} shows detailed information for the P(10) $v_{1-0}$ and P(4) $v_{2-1}$ line from escape probability for three representative TTauri disk models with $R_{\rm in}\!=\!0.1,1$ and 10~au. The three panels for each case show the optical depth in the respective line and continuum as a function of radius, the cumulative flux in the line and the line emitting region (15 and 85\% of the radial and vertical line flux respectively) on top of the CO density in the disk model.

\newpage

\section{Stellar and disk model parameters}

Table~\ref{CentralStars} shows the central star properties for our TTauri and Herbig star models, compared with \citet{bosman} DALI model parameters. In the same Table, we show the properties of our disk models that we keep fixed along all the model series. 

\begin{table*}
\centering
\caption{Overview of the fixed model parameters for the T~Tauri and Herbig models}
\begin{tabular}{p{4.5cm}p{2.5cm}p{2.5cm}p{2.5cm}p{2.5cm}}
\hline\hline
\multicolumn{5}{c}{Central star and radiation field parameters}\\
\hline
Parameter & Symbol & Value TTauri & Value Herbig & DALI, \citet{bosman} \\ \hline
Photospheric temperature & $T_\mathrm{eff}$ [K] & 4400 & 8600 & 10000\\
Stellar mass & $M_\mathrm{*}$ [M$_\mathrm{\odot}$] & 0.8 & 2.2 & 2.5\\
Stellar luminosity & $L_\mathrm{*}$ [L$_\mathrm{\odot}$] & 0.7 & 32 & 30\\
FUV excess & $L_\mathrm{UV}$/$L_\mathrm{*}$ & 0.01 & - & -\\
UV powerlaw exponent & $p_\mathrm{UV}$ & 0.2 & - & - \\
X-ray luminosity & $L_\mathrm{X}$ [erg/s] & 10$^{30}$ & - & - \\
X-ray minimum energy & $E_\mathrm{min,X}$ [keV] & 0.1 & - & -\\
X-ray Temperature & $T_\mathrm{X}$ [K] & 10$^7$ & - & -\\ \hline\hline
\multicolumn{4}{c}{Disk parameters of the standard model fixed in the
series}\\
\hline
Parameter & Symbol & ProDiMo & DALI\\ \hline
Radial $\times$ vertical grid points & $N_\mathrm{xx}\times\! N_\mathrm{zz}$ & 70 $\times$ 70 & -\\
Outer radius & $R_\mathrm{out}$ [au] & 300 & 500\\
Minimum dust size & $a_\mathrm{min}$ [$\mu$m] & 0.05 & -\\
Maximum dust size & $a_\mathrm{max}$ [mm] & 1 & -\\
Power law index of the dust size distribution & $a_\mathrm{pow}$ & 3.5 & - \\
Dust composition & - & Draine Astrosilicates$^{1}$ & - \\
Reference radius & $R_\mathrm{0}$ [au] & 50 & 50 \\
Scale height at reference radius & $H_\mathrm{0}$ [au] &
4.57 & 5\\
Scale height power law index & $\beta$ & 1.13 & 1.25\\
Tapering-off radius & $R_\mathrm{taper}$ [au] & 200 & 50\\
Surface density at $R_\mathrm{0}$ & $\Sigma_\mathrm{gas}(R_{0})$ [g/cm$^2$] & 1.0 & 60.0 \\
Chemical heating efficiency & - & 0.2 & -\\
Settling description & - & Dubrulle & parametrized\\
Cosmic ray ionization rate & $\zeta_\mathrm{CRs}$ [s$^{-1}$] &
1.7$\times10^{-17}$ & -\\
Distance & $d$ [pc] & 140 & 150\\
Turbulence viscosity coefficient & $\alpha_\mathrm{vis}$ & 0.05 & -\\
Disk inclination & [$\degr$] & 30 & 45\\
\hline
\end{tabular}
\tablefoot{$^1$ \citet{draine}}
\label{CentralStars}
\end{table*}


\end{document}